\documentclass[journal,twoside]{IEEEtran}
\usepackage{amsmath,amsfonts,amsthm,amssymb}
\usepackage{cite}
\usepackage{enumerate}
\usepackage{float}
\usepackage[keeplastbox]{flushend}
\usepackage{framed}
\usepackage{graphicx}
\usepackage{hyperref}
\usepackage[utf8]{inputenc}
\usepackage{mathrsfs}
\usepackage{mathtools}
\usepackage{subcaption}
\usepackage[dvipsnames]{xcolor}

\hypersetup{colorlinks=true,citecolor=black,linkcolor=black,pdfauthor=author}
\DeclareMathAlphabet{\mathpzc}{OT1}{pzc}{m}{it}
\newcommand{\orcid}[1]{\href{https://orcid.org/#1}{\includegraphics[width=7pt]{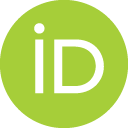}}}
\newtheorem{theorem}{Theorem}
\newtheorem{remark}{Remark}

\newtheorem{definition}{Definition}
\newtheorem{proposition}{Proposition}

\renewcommand{\theequation}{\arabic{equation}}

\begin{document}

\title{Underactuated Source Seeking by Surge Force Tuning: Theory and Boat Experiments}

\author{Bo~Wang\textsuperscript{\orcid{0000-0001-6047-1400}},~\IEEEmembership{Graduate Student Member,~IEEE,}
        Sergey~Nersesov\textsuperscript{\orcid{0000-0003-0331-1181}},~\IEEEmembership{Member,~IEEE,}
        Hashem~Ashrafiuon\textsuperscript{\orcid{0000-0001-5067-9064}},~\IEEEmembership{Senior~Member,~IEEE,}
        Peiman~Naseradinmousavi\textsuperscript{\orcid{0000-0002-0724-4070}},
        and~Miroslav~Krsti{\'c}\textsuperscript{\orcid{0000-0002-5523-941X}},~\IEEEmembership{Fellow,~IEEE}
\thanks{This research was supported in part by the U.S. Office of Naval Research under Grant N00014-19-1-2255. \textit{(Corresponding author: Hashem~Ashrafiuon.)}}   
\thanks{Bo Wang, Sergey Nersesov, and Hashem Ashrafiuon are with the Department of Mechanical Engineering, Villanova University, Villanova, PA 19085 USA (e-mail: bwang6@villanova.edu; sergey.nersesov@villanova.edu; hashem.ashrafiuon@ villanova.edu).}
\thanks{Peiman Naseradinmousavi is with the Department of Mechanical Engineering, San Diego State University, San Diego, CA 92182 USA (e-mail: pnaseradinmousavi@sdsu.edu).}
\thanks{Miroslav Krsti{\'c} is with the Department of Mechanical and Aerospace Engineering, University of California, San Diego, La Jolla, CA 92093 USA (e-mail: krstic@ucsd.edu).}
}

\markboth{IEEE TRANSACTIONS ON CONTROL SYSTEMS TECHNOLOGY}{Wang \MakeLowercase{\textit{et al.}}: Underactuated Source Seeking by Surge Force Tuning: Theory and Boat Experiments}

\maketitle

\begin{abstract}
We extend source seeking algorithms, in the absence of position and velocity measurements, and with the tuning of the surge input, from velocity-actuated (unicycle) kinematic models to force-actuated generic Euler-Lagrange dynamic underactuated models. In the design and analysis, we employ a symmetric product approximation, averaging, passivity, and partial-state stability theory. The proposed control law requires only real-time measurement of the source signal at the current position of the vehicle and ensures semi-global practical uniform asymptotic stability (SPUAS) with respect to the linear motion coordinates for the closed-loop system. The performance of our source seeker with surge force tuning is illustrated with both numerical simulations and experiments of an underactuated boat.
\end{abstract}

\begin{IEEEkeywords}
Extremum seeking, symmetric product approximation, planar underactuated vehicles, averaging, partial-state stability.
\end{IEEEkeywords}

\section{Introduction}

\subsection{Motivation}
Extremum seeking (ES) is a real-time model-free optimization approach that is applicable not only to static maps but also, somewhat uniquely, to dynamical systems \cite{ariyur2003real}. Following the development of the ES convergence guarantees by  \cite{krstic2000stability}, and their semi-global extension by \cite{tan2006non}, ES has been a flourishing research area, especially in the domain of autonomous vehicle control for finding sources of signals (electromagnetic, optical, chemical, etc.), distance-based localization, distance-based formation control, etc. The motivation for source seeking algorithms by and large comes from the fact that Global Positioning System (GPS) signals are not available in unstructured environments. Besides, the cost, weight, and complexity of onboard inertial navigation systems (INS) that do not drift over longer periods of time are prohibitive. Hence, autonomous vehicles that operate without GPS or INS benefit from source seeking capabilities.

Most real vehicles are \textit{underactuated}, whereby underactuated it is commonly meant that the number of independent actuators of a vehicle is strictly lower than the number of its degrees of freedom (DOF), as defined by the dimension of the configuration space \cite{bullo2005geometric}. As a consequence of the underactuation, the control design for these vehicles is much more difficult than for fully-actuated vehicles \cite{wang2021leader}. Specifically, 
fully-actuated {\em mechanical system} models (comprising the kinematic and dynamic equations) can be feedback linearized into double-integrator dynamics. This is not possible for underactuated vehicles. Furthermore, unlike (first-order) nonholonomic systems, where nonintegrable constraints are imposed on system velocities (such as in the unicycle), underactuated dynamic vehicle models describe the motions constrained by nonintegrable acceleration constraints, and thus, ES algorithms developed for first-order systems cannot be directly applied to underactuated vehicles.

\subsection{Related Results}
Given the rich variety of model types, spatial dimensions, and input tuning options for autonomous vehicles, numerous approaches for source seeking have emerged in the literature. We categorize the existing results into classical averaging-based, Lie bracket averaging-based, and symmetric product approximation-based seekers according to different types of averaging techniques.

\textit{Classical Averaging-Based Seekers.}
In \cite{zhang2007extremum}, two source seeking schemes were proposed for vehicles modeled as single and double integrators based on periodic averaging theory. The first source seeking method for the (velocity-controlled) unicycle model was presented in \cite{zhang2007source}, where the forward/surge velocity is tuned while the angular velocity is kept at a constant nonzero value (i.e., the vehicle is perpetually turning). The second result for unicycle source seeking, as an alternative to the surge velocity tuning, was the algorithm in \cite{cochran2009nonholonomic}, which keeps the forward speed constant while tuning the angular velocity. 
The method was later modified in \cite{ghods2010speed} to improve the performance, where the unicycle is allowed to slow down as it gets close to the source by regulating the forward speed. Instead of using periodic perturbations, discrete-time stochastic ES control laws were proposed in \cite{stankovic2009discrete,manzie2009extremum} for integrator systems. A further major shift from deterministic to stochastic approaches took place with the continuous-time source seeking algorithm by  \cite{liu2010stochastic}, which replaces sinusoidal probing with suitably filtered white noise, and where the stability analysis is conducted by novel stochastic averaging \cite{5415532}. In \cite{matveev2011navigation}, a sliding mode source seeking strategy was developed for velocity-controlled unicycles using the time derivative of the source measurements. 

\textit{Lie Bracket Averaging-Based  Seekers.}
A direction distinct from those employing classical averaging (periodic or stochastic)---such as in \cite{zhang2007extremum,zhang2007source,cochran2009nonholonomic,ghods2010speed,liu2010stochastic}---was charted by D{\"u}rr et al. \cite{durr2013lie} who introduced a Lie bracket averaging approach. The Lie bracket averaging-based strategies were applied to source seeking of single-integrator dynamics in \cite{durr2013lie} and velocity-controlled unicycles in \cite{durr2013lie,scheinker2014extremum,durr2017extremum}, where the semi-global practical stability of the source seeking systems is guaranteed. In \cite{suttner2017exponential,suttner2018formation}, the Lie bracket averaging-based ES approach was used in distance-based formation control for unicycles. Later, the Lie bracket averaging approach was also applied to the $n$-th order integrator-chain dynamics in \cite{michalowsky2014multidimensional,michalowsky2015model}. Although the methods proposed in \cite{zhang2007extremum,michalowsky2014multidimensional,michalowsky2015model} can be applied to force-controlled vehicles, they all depend on certain time-varying state transformations, and the stability analysis is based on the transformed system. As a consequence, these methods require the initial velocity of the vehicle to be larger than a certain constant, which depends on the perturbation frequency. Thus, these algorithms may not guarantee convergence when the vehicle starts from rest. This is the case also with the control law in \cite{scheinker2018extremum}, where the initial angular velocity is equal to the perturbation frequency. The Lie bracket averaging approach was generalized in \cite{grushkovskaya2018class}, where a broad class of control functions was presented for ES control. An adaptive ES scheme was proposed in \cite{suttner2019extremumseeking}, where the perturbation frequency is adaptively chosen such that the state trajectories exactly converge to the global minimum.

\textit{Symmetric Product Approximation-Based Seekers.}
The classical averaging methods and the Lie bracket averaging approaches cannot be applied directly to a generic second-order (force-controlled) vehicle model---Section 3 in \cite{zhang2007extremum} illustrates the need for additional compensation and analysis but applies only to a fully-actuated vehicle. The {\em symmetric product approximation} approach, which Bullo et al. \cite{bullo2000controllability,bullo2002averaging} introduced for vibrational control of mechanical systems, has enabled considerable further advances in force-actuated source seeking. The symmetric product approximation was first employed in source seeking with a force-controlled unicycle in \cite{suttner2019extremum} but assuming the availability of velocity measurements. The requirement of velocity measurements was removed by  \cite{suttner2020acceleration}. In these two papers, the angular motion dynamics of the unicycle are assumed to be a second-order integrator. The surge force is tuned by the source seeking algorithm, while the yaw torque is set to zero or to be periodic such that the orientation of the unicycle is a linear or periodic function of time. While these innovative works are the first to employ symmetric product approximation for source seeking of force-controlled vehicles, their model of rotational motion is simplified. In this paper, we conduct a design for a suitably modeled underactuated force-controlled vehicle. In a recent alternative pursuit by Suttner \cite{suttner2022extremumTAC} for fully-actuated mechanical systems with strict velocity-dependent dissipation, a symmetric product approximation-based ES controller was proposed and semi-global practical uniform asymptotic stability (SPUAS) was proved for the closed-loop system. In \cite{suttner2022extremum}, the symmetric product approach to ES control was extended to fully-actuated dissipation-free mechanical systems, where a phase-lead compensator injects damping into the system to achieve convergence. 
In addition, the method applies to systems on Lie groups including two- and three-dimensional vehicle models.

\subsection{Main Contributions}
In this paper, we develop a novel source seeking strategy for generic force-controlled planar underactuated vehicles.
The main contributions of this work are summarized as follows:
\begin{enumerate}[1)]
	\item We provide a theoretical foundation for ES algorithms based on symmetric product approximations. We prove that the trajectories of a class of underactuated mechanical systems can be approximated by the trajectories of corresponding \textit{symmetric product systems}. By incorporating symmetric product approximation, averaging, passivity, and partial-state stability theory, we show that the partial-state semi-global practical uniform asymptotic stability (P-SPUAS) of a class of underactuated mechanical systems follows from partial-state uniform global asymptotic stability (P-UGAS) of the corresponding symmetric product system.

    \item We consider the dynamic model of planar vehicles, instead of considering only the kinematic model such as in \cite{zhang2007source,cochran2009nonholonomic,scheinker2014extremum,durr2017extremum}. Furthermore, unlike the strategies presented in \cite{zhang2007extremum,michalowsky2014multidimensional,michalowsky2015model,scheinker2018extremum,suttner2022extremum} for fully-actuated vehicles, the proposed approach applies to strictly dissipative \textit{underactuated} vehicles, including boats/ships, planar underwater vehicles, etc. and allows the vehicle to start from rest if desired. 

    \item The presented seeking scheme does not require any position or velocity measurements. It requires only real-time measurements of the source signal at the current position of the vehicle and ensures SPUAS with respect to the linear motion coordinates for the closed-loop systems. The structure of the proposed controller is exceptionally simple and easy to implement: the measured output is multiplied by a periodic signal and fed into the surge force. 

\end{enumerate}

\subsection{Outline}
The paper is organized as follows. Preliminaries and problem formulation are given in Section \ref{sec:preliminaries}. Section \ref{sec:approximations} presents results on symmetric product approximations. In Section \ref{sec:controldesign}, we present the source seeking design and stability analysis for planar underactuated vehicles. Simulation results are shown in Section \ref{sec:simulations}. Experimental results are presented in Section \ref{sec:experiments}. Concluding remarks are provided in Section \ref{sec:conclusions}. The Appendices contain auxiliary results and proofs.

\subsection{Notation}
Let $\mathbb{R}^n$ denote the $n$-dimensional real vector space; $\mathbb{R}_{\ge 0}$ the set of all non-negative real numbers; $|\cdot|$ the Euclidean norm of vectors in $\mathbb{R}^n$. 
The gradient of a continuously differentiable function $f:\mathbb{R}^n\to\mathbb{R}$ is denoted by $\nabla f(x)\coloneqq\left[\frac{\partial f(x)}{\partial x_1},\ldots,\frac{\partial f(x)}{\partial x_n}\right]^\top$. 
For real matrices $A\in\mathbb{R}^{n\times m}$, we use the matrix norm $||A||=\sup\{|Ax|:|x|=1\}$.
For any constant $r>0$, we use the notation $\bar{\mathcal{B}}_r^n\coloneqq\{x\in\mathbb{R}^n:|x|\le r\}$ to denote a ball of radius $r$ in $\mathbb{R}^{n}$. 
For two vector fields $f,g:\mathbb{R}\times\mathbb{R}^n\to\mathbb{R}^n$, the Lie bracket is denoted by $(\operatorname{ad}_{g}f)(t,x)=[g,f](t,x)\coloneqq\frac{\partial f(t,x)}{\partial x}g(t,x)-\frac{\partial g(t,x)}{\partial x}f(t,x)$, and $\operatorname{ad}_{g}^k f\coloneqq\operatorname{ad}_{g}^{k-1} (\operatorname{ad}_{g}f)$. Throughout this paper, we omit the arguments of functions when they are clear from the context.

\section{Problem Statement}\label{sec:preliminaries}

\subsection{Model of Planar Underactuated Vehicles}
\begin{figure}[t]
	\centering
	\includegraphics[scale=0.5]{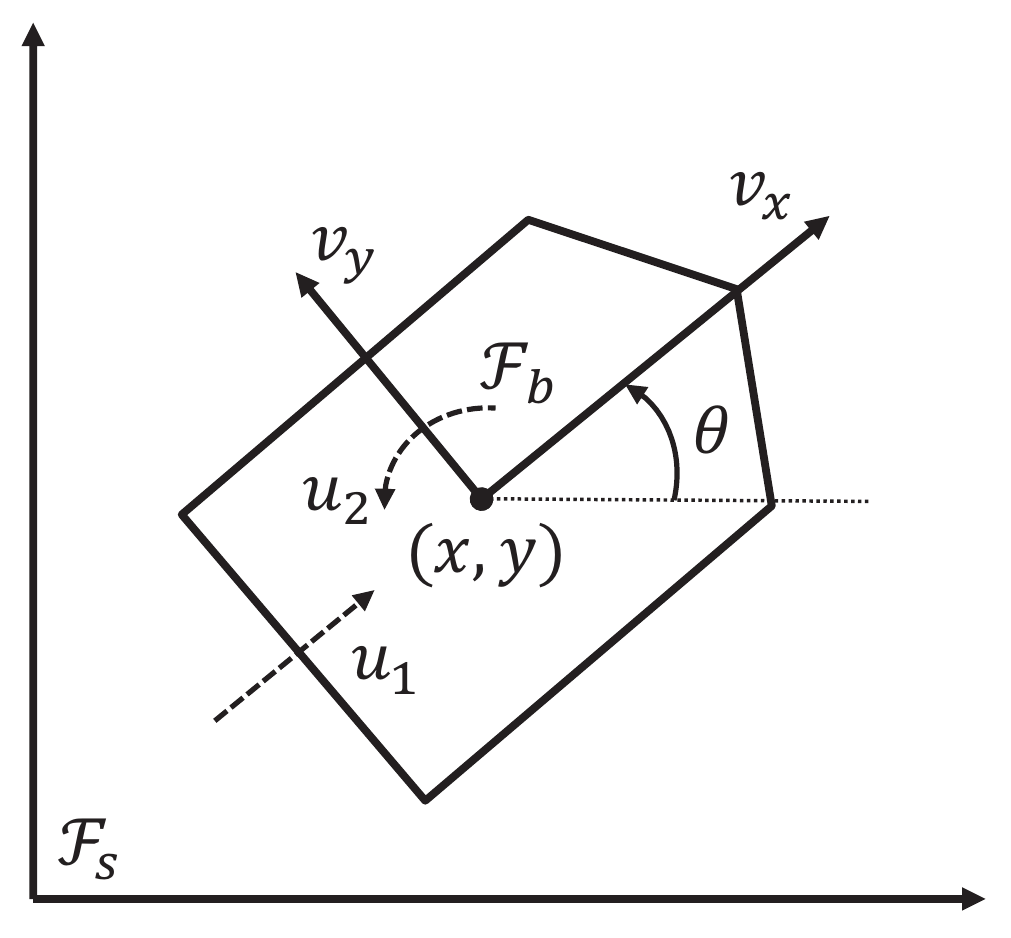}
	\caption{Top view of the planar underactuated vehicle.}
	\label{fig:vehicles}
\end{figure}

A generic planar underactuated vehicle can be modeled as a 3-DOF planar rigid body with two independent control inputs. Let $\mathcal{F}_s$ denote the fixed inertial frame attached to the ground, and $\mathcal{F}_b$ the body-fixed frame attached to the center of mass of the vehicle. The position of the vehicle in $\mathcal{F}_s$ is described by $(x,y)$, and the orientation of the vehicle is represented by $\theta$, as shown in Fig \ref{fig:vehicles}.
The equations of motion of the planar underactuated vehicle are given by
\begin{gather}
\dot{q}=J(q)v,\tag{1a}\label{eqn:lagrange-kinematics}\\
M\dot{v}+C(v)v+Dv=Gu \tag{1b}\label{eqn:lagrange-dynamics},
\end{gather}\setcounter{equation}{1}\noindent
where $q=[x,y,\theta]^\top\in\mathbb{R}^3$ is the configuration of the vehicle; $v=[v_x,v_y,\omega]^\top\in\mathbb{R}^3$ is the generalized velocity vector consisting of the linear velocity $(v_x, v_y)$ in the body-fixed frame and the angular velocity $\omega$; $u=[u_1,u_2]^\top\in\mathbb{R}^2$ is the control input vector; $J(q)$ is the kinematic transformation matrix given by
\begin{equation}
J(q)=\begin{bmatrix}
\cos(\theta) & -\sin(\theta) & 0\\
\sin(\theta) & \cos(\theta) & 0\\
0 & 0 & 1
\end{bmatrix};
\end{equation}
$M=\operatorname{diag\,}\{m_{11},m_{22},m_{33}\}>0$ is the inertia matrix; $C(v)=-C(v)^\top$ is the Coriolis matrix. The components of vector $C(v)v$ are homogeneous polynomials in $\{v_x,v_y,\omega\}$ of degree 2 \cite{bullo2005geometric}. We assume that the damping matrix $D$ is positive definite and constant, which implies that the damping force is proportional to the velocity. We also assume that the surge force and the yaw torque are the two independent control inputs, and accordingly, the input matrix $G$ is given by
\begin{equation}
G=\begin{bmatrix}
1 & 0\\
0 & 0\\
0 & 1
\end{bmatrix}.
\end{equation}
The system (\ref{eqn:lagrange-kinematics})-(\ref{eqn:lagrange-dynamics}) can model a wide class of planar underactuated vehicles such as ships, planar underwater vehicles, etc. 

\subsection{Control/Optimization Objective}
Assume that the position-dependent nonlinear cost function $\rho:\mathbb{R}^2\to\mathbb{R}_{\ge 0}$ is smooth and has a global extremum, i.e., there exists a unique $(x^\star,y^\star)\in\mathbb{R}^2$ such that
\begin{equation}\label{eqn:rho}
    \nabla\rho(x^\star,y^\star)=0~ {\rm and~}   \nabla\rho(x,y)\ne0, \forall (x,y)\ne (x^\star,y^\star).
\end{equation}
In applications, $\rho(\cdot)$ may represent the distance between the vehicle and a source, the strength of a certain (electromagnetic, optical, etc.) signal, or the concentration of chemical materials. Without loss of generality, we assume that  $(x^\star,y^\star)$ is the minimum of the function $\rho$ and the vehicle can measure the value of $\rho(x(t),y(t))$ in real-time. Note that both the extremum $(x^\star,y^\star)$ and the gradient $\nabla\rho$ are unknown. Given any constant $\varepsilon>0$, the objective is to develop a feedback controller to steer the vehicle  without position and velocity measurements such that
\begin{equation}
    \lim_{t\to\infty}|(x(t),y(t))-(x^\star,y^\star)|\le \varepsilon.
\end{equation}

\subsection{Shifted Passivity}

In the existing literature, there are generally two types of source seeking schemes for vehicle systems: 1) tuning the forward motion of the vehicle by the ES loop while keeping the angular speed constant (e.g., \cite{zhang2007source,suttner2019extremum,durr2013lie}), and 2) tuning the angular motion of the vehicle by the ES loop while keeping the forward velocity constant (e.g., \cite{cochran2009nonholonomic,liu2010stochastic,durr2017extremum}). In either case, the desired (linear/angular) velocity component is not zero, but instead has a steady-state value corresponding to a non-zero constant input.
We formulate this property from the viewpoint of \textit{shifted passivity} \cite{monshizadeh2019conditions}. 

Consider the system (\ref{eqn:lagrange-kinematics})-(\ref{eqn:lagrange-dynamics}) with the output $\eta\coloneqq G^\top v$. Define the steady-state set
\begin{equation}
\mathcal{E}\coloneqq \{(v,u)\in\mathbb{R}^3\times \mathbb{R}^2:C(v)v+Dv-Gu=0  \}.
\end{equation}
Fix $(v^*,u^*)\in\mathcal{E}$ and the steady-state output $\eta^* \coloneqq G^\top v^*$. 

\begin{definition}[Shifted passivity]\rm
The system (\ref{eqn:lagrange-kinematics})-(\ref{eqn:lagrange-dynamics}) is said to be \textit{shifted passive} if the input-output mapping $(u-u^*)\mapsto (\eta-\eta^*)$ is passive, i.e., there exists a storage function $\mathcal{H}:\mathbb{R}^{3}\to\mathbb{R}_{\ge 0}$ such that for all $(v,u)\in\mathbb{R}^3\times \mathbb{R}^2$,
\begin{equation}
\dot{\mathcal{H}}\coloneqq (\nabla \mathcal{H}(v))^\top\dot{v}\le (u-u^*)^\top(\eta-\eta^*).
\end{equation}
\end{definition}
 
\begin{proposition}\label{prop:shifted-passivity} 
Consider the system (\ref{eqn:lagrange-kinematics})-(\ref{eqn:lagrange-dynamics}) with the steady-state input $u^*=[0,c]^\top$, where $c>0$ is a constant. Then, there exists $\hat{c}>0$ such that for all $c\in(0,\hat{c})$, the system (\ref{eqn:lagrange-kinematics})-(\ref{eqn:lagrange-dynamics}) is shifted passive.
\end{proposition}
\begin{proof}
Fix the input $u^*=[0,c]^\top$, and the corresponding steady-state velocity and output are $v^*=[0,0,\omega^*]^\top$ and $\eta^*=[0,\omega^*]^\top$, respectively.
Let the storage function be $\mathcal{H}(v)=\frac{1}{2}(v-v^*)^\top M(v-v^*)$. Then, the time derivative of $\mathcal{H}(v)$ along the trajectories of (\ref{eqn:lagrange-kinematics})-(\ref{eqn:lagrange-dynamics}) is given by
\begin{align}
\dot{\mathcal{H}}&= (v-v^*)^\top \left[G(u-u^*)-C(v)v-Dv +Gu^* \right] \notag\\
&=(\eta-\eta^*)^\top(u-u^*)  - (v-v^*)^\top\left[C(v)v+Dv-Gu^*  \right] \notag\\
&= (\eta-\eta^*)^\top(u-u^*)-(v-v^*)^\top D(v-v^*) \notag\\
&\quad -(v-v^*)^\top\left[C(v)-C(v^*)\right]v^* \label{eqn:dot-H},
\end{align}
where we used (\ref{eqn:lagrange-dynamics}), and added and subtracted the term $Gu^*$ in the first identity, added and subtracted the term $(C(v)+D)v^*$ in the second identity, and used $Gu^*=C(v^*)v^*+Dv^*$ and the skew-symmetric property of $C(v)$ in the third identity. Let us denote $\mathcal{J}(v)\coloneqq C(v)v^*+Dv$.
From (\ref{eqn:dot-H}) we have
\begin{equation}
\dot{\mathcal{H}}= (\eta-\eta^*)^\top(u-u^*) - (v-v^*)^\top \left[\mathcal{J}(v)-\mathcal{J}(v^*)\right].
\end{equation}
It follows from the homogeneity of $C(v)v$ that for all $v\in\mathbb{R}^3$, $||\partial \left[C(v)e_3 \right]/\partial v ||$ is bounded, where $e_3=[0,0,1]^\top$. Thus, we can always choose $\omega^*$ small enough such that $\partial \left[C(v)v^* \right]/\partial v + \left[\partial \left[C(v)v^* \right]/\partial v\right]^\top \le 2D$, which implies that $(\partial \mathcal{J}(v)/\partial v)+(\partial \mathcal{J}(v)/\partial v)^\top\ge 0$ for all $v\in\mathbb{R}^3$. Therefore, the map $\mathcal{J}(\cdot)$ is monotone, and correspondingly, $(v-v^*)^\top \left[\mathcal{J}(v)-\mathcal{J}(v^*)\right]\ge 0$, which completes the proof.
\end{proof}

\subsection{Partial-State Practical Stability}
Consider the nonlinear interconnected system
\begin{align}
    \dot{x}_1 &= f_1(x_1,x_2), \quad x_1(t_0)=x_{10}, \quad t\ge t_0, \label{eqn:interconnected-1}\\
    \dot{x}_2 &= f_2(x_1,x_2), \quad x_2(t_0)=x_{20},\label{eqn:interconnected-2}
\end{align}
where $f_1:\mathbb{R}^{n_1}\times \mathbb{R}^{n_2}\to \mathbb{R}^{n_1}$ is such that, for every $x_2\in\mathbb{R}^{n_2}$, $f_1(0,x_2)=0$ and $f_1(x_1,x_2)$ is locally Lipschitz in $x_1$ uniformly in $x_2$; $f_2:\mathbb{R}^{n_1}\times \mathbb{R}^{n_2}\to \mathbb{R}^{n_2}$ is such that for every $x_1\in\mathbb{R}^{n_1}$, $f_2(x_1,x_2)$ is locally Lipschitz in $x_2$ uniformly in $x_1$. Let $x_1(\cdot)\coloneqq x_1(\cdot,x_{10},x_{20})$ and $x_2(\cdot)\coloneqq x_2(\cdot,x_{10},x_{20})$ denote the solution of the initial value problem (\ref{eqn:interconnected-1})-(\ref{eqn:interconnected-2}).  We define the \textit{partial-state stability} as stability with respect to $x_1$ for system (\ref{eqn:interconnected-1})-(\ref{eqn:interconnected-2}).

\begin{definition}[P-UGAS]\rm \label{def:partial-stability}
The system (\ref{eqn:interconnected-1})-(\ref{eqn:interconnected-2}) is \textit{globally asymptotically stable (GAS) with respect to $x_1$ uniformly in $x_{20}$} if the following conditions are satisfied:
\begin{enumerate}[1)]\setlength{\itemsep}{-0.1cm}
    \item \textit{Partial-State Uniform Stability (P-US):} For each $\varepsilon>0$, there exists $\delta(\varepsilon)$ such that 
    \begin{equation*}
    |x_{10}|\le\delta(\varepsilon) \implies |x_1(t)|\le \varepsilon, \quad \forall t\ge 0,~ \forall x_{20}\in\mathbb{R}^{n_2}.
    \end{equation*}
    \item \textit{Partial-State Uniform Global Boundedness (P-UGB):} For each $r>0$, there exists $R(r)$ such that
    \begin{equation*}
    |x_{10}|\le r \implies |x_1(t)|\le R(r), \quad \forall t\ge 0,~ \forall x_{20}\in\mathbb{R}^{n_2}.
\end{equation*}
    \item \textit{Partial-State Uniform Global Attractivity (P-UGA):} For each $r>0$, for each $\sigma>0$, there exists $T(r,\sigma)$ such that 
\begin{equation*}
    |x_{10}|\le r \implies |x_1(t)|\le \sigma, \quad \forall t\ge T(r,\sigma),~ \forall x_{20}\in\mathbb{R}^{n_2}.
\end{equation*}
\end{enumerate}
\end{definition}

The partial-state stability, which is also referred to as ``partial stability” in the literature \cite{hahn1967stability,grushkovskaya2019partial,wang2021use}, is a special case of ``output stability" \cite{karafyllis2021lyapunov,sontag2000lyapunov,teel2000smooth}. By viewing $x_1$ as the output, the P-UGAS in Definition \ref{def:partial-stability} is equivalent to the uniform global asymptotic output stability (UGAOS)  \cite[Definition 1]{karafyllis2021lyapunov} for (\ref{eqn:interconnected-1})-(\ref{eqn:interconnected-2}).

We present Lyapunov conditions for P-UG(A)S of (\ref{eqn:interconnected-1})-(\ref{eqn:interconnected-2}). Given a function $V(x_1,x_2)$, define $\dot{V}(x_1,x_2)= (\partial V/\partial x)f(x_1,x_2)$, where $x=[x_1^\top,x_2^\top]^\top$ and $f(x_1,x_2)= [f_1(x_1,x_2)^\top,f_2(x_1,x_2)^\top]^\top$.
\begin{theorem}[\cite{haddad2011nonlinear}]\label{thm:P-UGAS}
Consider the interconnected system (\ref{eqn:interconnected-1})-(\ref{eqn:interconnected-2}). 
If there exist a function $V:\mathbb{R}^{n_1}\times \mathbb{R}^{n_2}\to \mathbb{R}_{\ge 0}$ of class $C^1$, class-$\mathcal{K}_\infty$ functions $\alpha_1,\alpha_2$ such that for all $(x_1,x_2)\in\mathbb{R}^{n_1}\times \mathbb{R}^{n_2}$,
\begin{gather}
    \alpha_1(|x_1|)\le V(x_1,x_2) \le \alpha_2(|x_1|), \label{eqn:positive-decrescent}\\
    \dot{V}(x_1,x_2)\le 0,
\end{gather}
then the system (\ref{eqn:interconnected-1})-(\ref{eqn:interconnected-2}) is US and UGB with respect to $x_1$ uniformly in $x_{20}$. Furthermore, if exists a positive definite function $\alpha_3$ such that for all $(x_1,x_2)\in\mathbb{R}^{n_1}\times \mathbb{R}^{n_2}$,
\begin{equation}
    \dot{V}(x_1,x_2)\le -\alpha_3(|x_1|),
\end{equation}
then (\ref{eqn:interconnected-1})-(\ref{eqn:interconnected-2}) is UGAS with respect to $x_1$ uniformly in $x_{20}$.
\end{theorem}

Next, we define \textit{partial-state practical stability} for interconnected systems that depends on a small parameter $\varepsilon>0$,
\begin{align}
    \dot{x}_1 &= f^\varepsilon_1(t,x_1,x_2), \quad x^\varepsilon_1(t_0)=x_{10}, \quad t\ge t_0, \label{eqn:epsilon-interconnected-1}\\
    \dot{x}_2 &= f^\varepsilon_2(t,x_1,x_2). \quad x^\varepsilon_2(t_0)=x_{20},\label{eqn:epsilon-interconnected-2}
\end{align}
Let $x^\varepsilon_1(\cdot)\coloneqq x^\varepsilon_1(\cdot,t_0,x_{10},x_{20})$ and $x^\varepsilon_2(\cdot)\coloneqq x^\varepsilon_2(\cdot,t_0,x_{10},x_{20})$ denote the solution of the initial value problem (\ref{eqn:epsilon-interconnected-1})-(\ref{eqn:epsilon-interconnected-2}).

\begin{definition}[P-SPUAS]\rm \label{def:partial-practical-stability}
The system (\ref{eqn:epsilon-interconnected-1})-(\ref{eqn:epsilon-interconnected-2}) said to be \textit{semi-globally practically asymptotically stable (SPAS) with  respect  to $x_1$ uniformly in $(t_0,x_{20})$} if for every compact set $\bar{\mathcal{B}}_r^{n_2}\subset\mathbb{R}^{n_2}$, the following conditions are satisfied:
\begin{enumerate}[1)]\setlength{\itemsep}{-0.1cm}
    \item For every $c_2>0$, there exists $c_1$ and $\hat{\varepsilon}(r)>0$ such that for all $(t_0,x_{20})\in\mathbb{R}_{\ge 0}\times \bar{\mathcal{B}}_r^{n_2} $ and for all $\varepsilon\in (0,\hat{\varepsilon})$,
    \begin{equation*}
        |x_{10}|\le c_1 \implies |x^\varepsilon_1(t)|\le c_2, \quad \forall t\ge t_0.
    \end{equation*}
    \item For every $c_1>0$, there exists $c_2$ and $\hat{\varepsilon}(r)>0$ such that for all $(t_0,x_{20})\in\mathbb{R}_{\ge 0}\times \bar{\mathcal{B}}_r^{n_2}$ and for all $\varepsilon\in (0,\hat{\varepsilon})$,
    \begin{equation*}
        |x_{10}|\le c_1 \implies |x^\varepsilon_1(t)|\le c_2, \quad \forall t\ge t_0.
    \end{equation*}
    \item For all $c_1>0$, $c_2>0$, there exists $T(c_1,c_2)$ and $\hat{\varepsilon}(r)>0$ such that for all $(t_0,x_{20})\in\mathbb{R}_{\ge 0}\times \bar{\mathcal{B}}_r^{n_2} $ and for all $\varepsilon\in (0,\hat{\varepsilon})$,
    \begin{equation*}
        |x_{10}|\le c_1 \implies |x^\varepsilon_1(t)|\le c_2, \quad \forall t\ge t_0+T(c_1,c_2).
    \end{equation*}
\end{enumerate}
\end{definition}
The notion of P-SPUAS is an extension of the notion of SPUAS \cite{teel1999semi,durr2013lie,moreau2000practical}. It is well known that, under the assumption that trajectories of (\ref{eqn:epsilon-interconnected-1})-(\ref{eqn:epsilon-interconnected-2}) converge to trajectories of (\ref{eqn:interconnected-1})-(\ref{eqn:interconnected-2}) uniformly on compact time intervals as $\varepsilon\to 0$, if  (\ref{eqn:interconnected-1})-(\ref{eqn:interconnected-2}) is GAS, then the origin of  (\ref{eqn:epsilon-interconnected-1})-(\ref{eqn:epsilon-interconnected-2}) is SPUAS \cite{teel1999semi,moreau2000practical}. We extend this claim to interconnected systems with partial-state stability.

\begin{definition}[Partial Converging Trajectories Property]\rm
The systems (\ref{eqn:interconnected-1})-(\ref{eqn:interconnected-2}) and (\ref{eqn:epsilon-interconnected-1})-(\ref{eqn:epsilon-interconnected-2}) are said to satisfy the \textit{partial converging trajectories property} if for every $T>0$, for every compact set $K\subset\mathbb{R}^{n_1}\times \mathbb{R}^{n_2}$, and for every $d>0$, there exists $\varepsilon^*$ such that for all $t_0\ge 0$, for all $(x_{10},x_{20})\in K$ and for all $\varepsilon\in(0,\varepsilon^*)$,
\begin{equation}
    |x_1^\varepsilon(t)-x_1(t)|<d,\quad \forall t\in[t_0,t_0+T].
\end{equation}
\end{definition}

\begin{proposition}\label{thm:partial-SPUAS}
Assume that for the system (\ref{eqn:interconnected-1})-(\ref{eqn:interconnected-2}), the flow $(x_1(\cdot),x_2(\cdot))$ is forward complete, and that the systems (\ref{eqn:interconnected-1})-(\ref{eqn:interconnected-2}) and (\ref{eqn:epsilon-interconnected-1})-(\ref{eqn:epsilon-interconnected-2}) satisfy the partial converging trajectories property.
If (\ref{eqn:interconnected-1})-(\ref{eqn:interconnected-2}) is GAS with respect to $x_1$ uniformly in $x_{20}$, then (\ref{eqn:epsilon-interconnected-1})-(\ref{eqn:epsilon-interconnected-2}) is SPAS with respect to $x_1$ uniformly in $(t_0,x_{20})$.
\end{proposition}

The proof of Proposition \ref{thm:partial-SPUAS} is given in Appendix \ref{appendix:stability}.

\section{Symmetric Product Approximations}\label{sec:approximations}

\subsection{Motivational Example}
The classical averaging technique \cite{zhang2007extremum,zhang2007source} and the Lie bracket averaging approach \cite{durr2013lie} cannot be directly applied to the system (\ref{eqn:lagrange-kinematics})-(\ref{eqn:lagrange-dynamics}). The classical averaging technique applies to systems in the form
\begin{equation}\label{eqn:averaging-form}
    \dot{\xi}=\varepsilon f(t,\xi,\varepsilon),
\end{equation}
where $\varepsilon>0$ is a small parameter and $f$ is (almost) periodic in $t$.
However, it is not possible to find a transformation for rewriting  system (\ref{eqn:lagrange-kinematics})-(\ref{eqn:lagrange-dynamics}) into the form (\ref{eqn:averaging-form}) in general \cite{khalil2002nonlinear}. The Lie bracket averaging approach applies to input-affine systems in the form \cite{durr2013lie,grushkovskaya2018class}
\begin{equation}\label{eqn:lie-bracket-form}
    \dot{\xi}=b_0(t,\xi)+\sum_{i=1}^m b_i(t,\xi)\sqrt{\omega}u_i(t,\omega t),
\end{equation}
where $\omega\in(0,\infty)$, $m$ is a positive integer,
and the corresponding Lie bracket system is given by
\begin{equation}
    \dot{\zeta}=b_0(t,\zeta)+\sum_{\substack{i=1\\j=i+1}}^m[b_i,b_j](t,\zeta)w_{ji}(t),
\end{equation}
where $w_{ji}(t)=\frac{1}{T}\int_0^T u_j(t,s)\int_0^s u_i(t,\tau){\rm d}\tau {\rm d}s$.
For illustration, let us consider a damped double-integrator system
\begin{equation}\label{eqn:double-integrator}
\begin{bmatrix}
\dot{\xi}_1\\ \dot{\xi}_2
\end{bmatrix}=
\underbrace{\begin{bmatrix}
\xi_2\\ -\xi_2
\end{bmatrix}}_{b_0(\xi_2)}+\sum_{i=1}^m
\underbrace{\begin{bmatrix}
0\\ k_i(\xi_1)
\end{bmatrix}}_{b_i(\xi_1)}u_i(t).
\end{equation}
where $\xi_1,\xi_2,u_i\in\mathbb{R}$,  $\xi=[\xi_1,\xi_2]^\top$, and $k_i(\cdot)$'s represent arbitrary  $\xi_1$ (position)-dependent  functions. A simple calculation shows that the Lie brackets between any two input vector fields are zero, i.e., $[b_i,b_j]\equiv 0$ for any $i,j=1,\ldots,m$, and thus, the Lie bracket approximations cannot be applied to the double-integrator system (\ref{eqn:double-integrator}), let alone the system (\ref{eqn:lagrange-kinematics})-(\ref{eqn:lagrange-dynamics}).

Next, we will show that the symmetric product approximations can be used to solve the ES problem for (\ref{eqn:lagrange-kinematics})-(\ref{eqn:lagrange-dynamics}). To illustrate the main idea, consider (\ref{eqn:double-integrator}) again. 
We first change the time scale by setting $\tau=t/\varepsilon$, and let $u_i(t)=(1/\varepsilon)v_i(t/\varepsilon)$. Then, (\ref{eqn:double-integrator}) becomes
\begin{equation}\label{eqn:double-integrator1}
\frac{{\rm d}}{{\rm d}\tau} \xi=\varepsilon f(\xi)
+g(\tau,\xi),
\end{equation}
where $f(\xi)=b_0(\xi_2)$ and $g(\tau,\xi)=\sum_{i=1}^m b_i(\xi_1)v_i(\tau)$. According to the variation of constants formula given in Appendix \ref{appendix:geometric}, the corresponding \textit{pull back system} is given by
\begin{equation}\label{eqn:pull-back0}
    \frac{{\rm d}}{{\rm d}\tau}z=\varepsilon F(\tau,z), \quad z(0)=\xi(0),
\end{equation}
where $z=[z_1,z_2]^\top$ and
\begin{align*}
    &F(\tau,z)=f(z)\\
    &\, +\sum_{k=1}^\infty \int_0^\tau\cdots\int_0^{s_{k-1}}\left({\rm ad}_{g(s_k,z)}\cdots{\rm ad}_{g(s_1,z)}f(z)\right) {\rm d}s_k\cdots{\rm d}s_1.
\end{align*}
By direct calculations, we have
\begin{equation}\label{eqn:int1}
    {\rm ad}_{g(s_1,z)}f(z)=-\sum_{i=1}^m v_i(s_1)
    \begin{bmatrix}
    -k_i(z_1)\\
    k_i(z_1)+k_i'(z_1)z_2
    \end{bmatrix},
\end{equation}
and
\begin{equation}\label{eqn:int2}
    {\rm ad}_{g(s_2,z)}{\rm ad}_{g(s_1,z)}f(z)=-\sum_{i,j=1}^m v_i(s_1)v_j(s_2)
    \begin{bmatrix}
    0\\
    \left(k_i(z_1)k_j(z_1)\right)'
    \end{bmatrix}.
\end{equation}
Note that the structural property of the system (\ref{eqn:double-integrator}) guarantees that the higher order terms ${\rm ad}_{g(s_k,z)}\ldots{\rm ad}_{g(s_1,z)}f(z)\equiv 0$ for all $k\ge 3$. Thus, the pull back vector field $F$ can be written as
\begin{equation*}\label{eqn:pull-back1}
\begin{split}
    F&(\tau,z)=f(z)-\sum_{i=1}^m\begin{bmatrix}
    -k_i(z_1)\\
    k_i(z_1)+k_i'(z_1)z_2
    \end{bmatrix}\int_0^\tau v_i(s_1){\rm d}s_1\\
    &\quad -\sum_{i,j=1}^m\begin{bmatrix}
    0\\
    \left(k_i(z_1)k_j(z_1)\right)'
    \end{bmatrix}\int_0^\tau\int_0^{s_1} v_i(s_1)v_j(s_2){\rm d}s_2{\rm d}s_1.
\end{split}
\end{equation*}
Denote the solution of the pull back system (\ref{eqn:pull-back0}) by $z(t)$.
Then, it follows from the variation of constants formula in Appendix \ref{appendix:geometric} that the solution of the system (\ref{eqn:double-integrator1}) is given by the initial value problem 
\begin{equation}\label{eqn:ivp}
\frac{{\rm d}}{{\rm d}\tau}\begin{bmatrix}
{\xi}_1\\ {\xi}_2
\end{bmatrix}=\sum_{i=1}^m
\begin{bmatrix}
0\\  k_i(\xi_1)
\end{bmatrix}v_i(\tau), \quad\xi(0)=z(\tau).
\end{equation}
We change the time scale back to $t=\varepsilon \tau$, and it follows from (\ref{eqn:ivp}) that $\dot{\xi}_1\equiv 0$, which implies that $\xi_1(t)\equiv \xi_1(0)\equiv  z_1(t)$. That is, the position trajectory of the double-integrator system (\ref{eqn:double-integrator}) is the $z_1$-trajectory of the pull back system (\ref{eqn:pull-back0}).

\begin{figure}[t]
	\centering
	\includegraphics[scale=0.5]{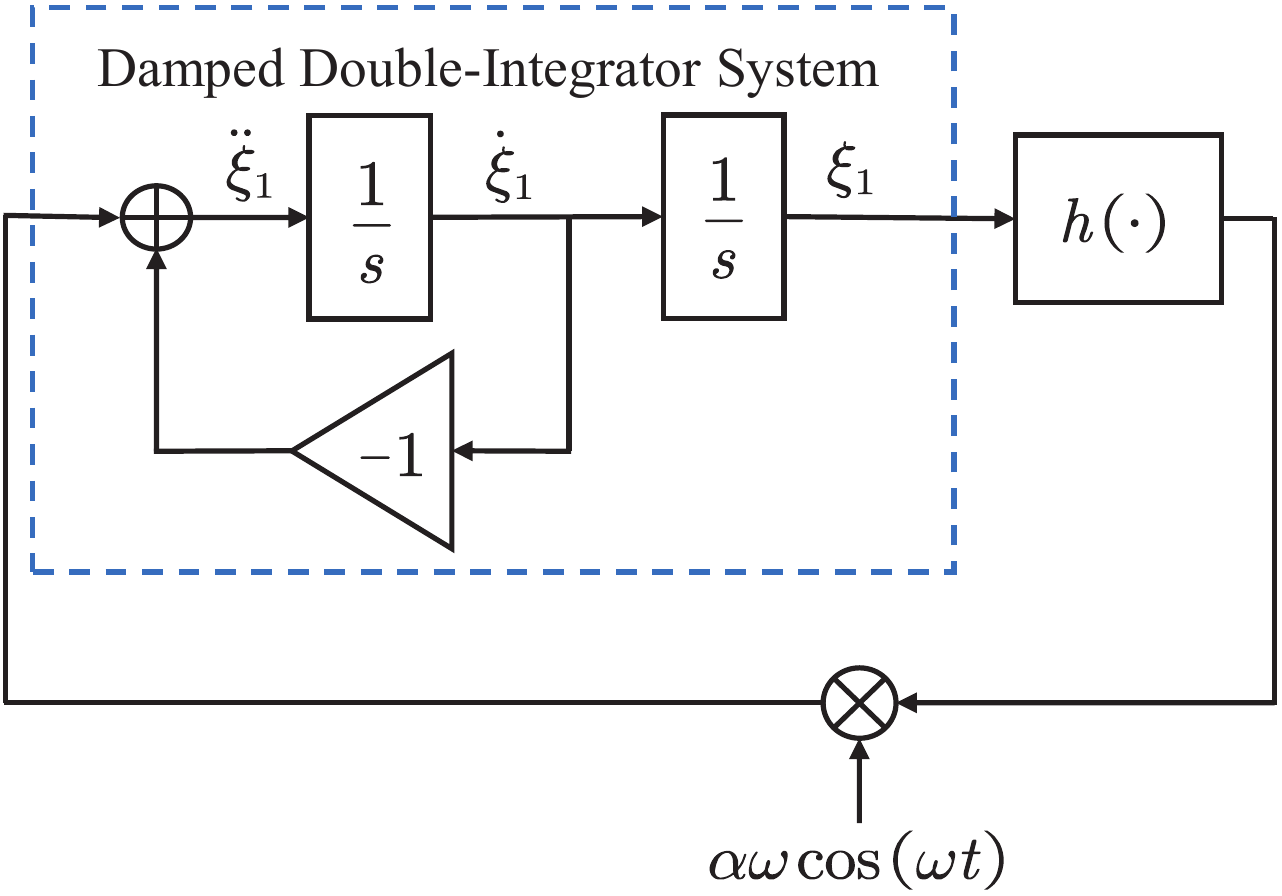}
	\caption{ES scheme for the damped double-integrator system.}
	\label{fig:second-order}
\end{figure}

The basic ES scheme for the double-integrator system  (\ref{eqn:double-integrator}) is illustrated in Fig. \ref{fig:second-order}. The closed-loop system can be written as 
\begin{equation}\label{eqn:double-integrator2}
\begin{bmatrix}
\dot{\xi}_1\\ \dot{\xi}_2
\end{bmatrix}=
\begin{bmatrix}
\xi_2\\ -\xi_2
\end{bmatrix}
+
\begin{bmatrix}
0\\  h(\xi_1) 
\end{bmatrix}\alpha\omega \cos(\omega t),
\end{equation}
where $h(\cdot)$ is the cost function and $\alpha\in\mathbb{R}$ is a constant.
The pull back system in time scale $\tau=t/\varepsilon=\omega t$ is given by
\begin{equation*}
    \begin{split}
    \frac{{\rm d}z}{{\rm d}\tau}=\varepsilon \left\{f(z)-\alpha\begin{bmatrix}
    -h(z_1)\\
    h(z_1)+h'(z_1)z_2
    \end{bmatrix}\sin(\tau) \right.\\
    \left.- \frac{\alpha^2}{2}\begin{bmatrix}
    0\\
    \left(h^2(z_1)\right)'
    \end{bmatrix}\sin^2(\tau)\right\}.
\end{split}
\end{equation*}
The pull back system is in the form of (\ref{eqn:averaging-form}), and the averaged pull back system in time scale $t$ is given by \cite[Section 10.4]{khalil2002nonlinear}
\begin{equation}\label{eqn:averaged}
    \begin{bmatrix}
\dot{\bar{z}}_1\\  \dot{\bar{z}}_2
\end{bmatrix}=\begin{bmatrix}
\bar{z}_2\\  -\bar{z}_2
\end{bmatrix} - \frac{\alpha^2}{4}\begin{bmatrix}
    0\\
    2h(\bar{z}_1)\nabla h(\bar{z}_1)
    \end{bmatrix}.
\end{equation}
Letting $V(\bar{z}_1,\bar{z}_2)=\frac{\alpha^2}{4}h^2(\bar{z}_1) + \frac{1}{2}\bar{z}_2^2$ and taking the time derivative along trajectories of (\ref{eqn:averaged}), we have $\dot{V}=-\bar{z}_2^2\le 0$. If the cost function $h(\cdot)$ has a global minimum $h(\xi_1^\star)\ge 0$, then it follows from the Krasovskii-LaSalle principle that the global minimum is GAS. Finally, using the averaging theorem in \cite{teel1999semi} and the conclusion that $\xi_1(t)\equiv z_1(t)$, the global minimum is SPUAS for the closed-loop system (\ref{eqn:double-integrator2}), i.e.,  the position trajectory $\xi_1(t)$ converges to an $O(\varepsilon)$-neighborhood of the global minimum point $\xi_1^\star$ as $t\to \infty$. We will now generalize this idea to system (\ref{eqn:lagrange-kinematics})-(\ref{eqn:lagrange-dynamics}).

\subsection{Symmetric Product Approximations}

Consider the system (\ref{eqn:lagrange-kinematics})-(\ref{eqn:lagrange-dynamics}). Let the input vector be
\begin{equation}\label{eqn:input}
    u=b_0+\frac{1}{\varepsilon}\sum_{i=1}^m b_i(q)w_i\left(\frac{t}{\varepsilon}\right),
\end{equation}
where  $\varepsilon$ is a positive constant, $m$ is a positive integer, $b_0=[b_{10},b_{20}]^\top$ is a constant vector, $b_i(q)=[b_{1i}(q),b_{2i}(q)]^\top$, and $\{w_i(t)\}$ are $T$-periodic functions satisfying
\begin{equation}\label{eqn:periodic-1}
    \int_0^T w_{i}(s_1){\rm d}s_1=0, \quad i=1,\ldots,m,
\end{equation}
\begin{equation}\label{eqn:periodic-2}
    \int_0^T \int_0^{s_2} w_{i}(s_1){\rm d}s_1{\rm d}s_2=0, \quad i=1,\ldots,m.
\end{equation}
Then, (\ref{eqn:lagrange-kinematics})-(\ref{eqn:lagrange-dynamics}) with the input vector (\ref{eqn:input}) in time scale  $\tau=t/\varepsilon$ can be written as
\begin{equation}\label{eqn:closed-loop}
    \frac{{\rm d}}{{\rm d}\tau}
    \begin{bmatrix}
    q\\v
    \end{bmatrix}=
    \varepsilon
    \underbrace{\begin{bmatrix}
    J(q)v\\-M^{-1}[C(v)v+Dv-B_0]
    \end{bmatrix}}_{\mathsf{f}(q,v)}+\underbrace{
    \begin{bmatrix}
    0\\ \sum\limits_{i=1}^{m}B_iw_i(\tau)
    \end{bmatrix}}_{\mathsf{g}(\tau,q)},
\end{equation}
where $B_0= Gb_0$ and $B_{i}(q)= M^{-1}G b_{i}(q)$ for $i=1,\ldots,m$.
Denote $\mathsf{f}_2(v)= -M^{-1}[C(v)v+Dv-B_0]$.
The \textit{symmetric product} of two vector fields $X,Y:\mathbb{R}^3\to\mathbb{R}^3$ corresponding to system (\ref{eqn:lagrange-kinematics})-(\ref{eqn:lagrange-dynamics}) is defined as
\begin{equation}\label{eqn:symmetric-product}
    \langle X:Y\rangle=\frac{\partial X}{\partial q}J(q)Y + \frac{\partial Y}{\partial q}J(q)X-\left(\frac{\partial}{\partial v}\left(\frac{\partial \mathsf{f}_2}{\partial v}X\right)\right)Y.
\end{equation}
The symmetric product $\langle\cdot:\cdot\rangle$ satisfies $\langle X:Y\rangle=\langle Y:X\rangle$.

In the next theorem, we show that system (\ref{eqn:lagrange-kinematics})-(\ref{eqn:lagrange-dynamics})  with input (\ref{eqn:input})  can be approximated by the \textit{symmetric product system}
\begin{align}
\dot{\bar{q}}&=J(\bar{q})\bar{v},\tag{34a}\label{eqn:lagrange-kinematics1}\\
M\dot{\bar{v}}+C(\bar{v})\bar{v}+D\bar{v}&=B_{0}-M \sum_{i,j=1}^m \Lambda_{ij}  \langle B_{i}:B_{j} \rangle(\bar{q}) \tag{34b}\label{eqn:lagrange-dynamics1},
\end{align}\setcounter{equation}{34}\noindent
where
\begin{equation}
    \Lambda_{ij}=\frac{1}{2T}\int_0^T \left(\int_0^{s_1}w_{i}(s_2) {\rm d}s_2\right)\left(\int_0^{s_1}w_{j}(s_2) {\rm d}s_2\right)  {\rm d}s_1.
\end{equation}
Define the time-varying vector field as
\begin{equation}
\Xi(t,q)\coloneqq \sum_{i=1}^m \left(\int_{0}^{t}w_i(s){\rm d}s\right)B_i(q).
\end{equation}

\begin{theorem}\label{thm:symmetric-product}
Consider the system (\ref{eqn:lagrange-kinematics})-(\ref{eqn:lagrange-dynamics})  with input vector (\ref{eqn:input}) and the symmetric product system (\ref{eqn:lagrange-kinematics1})-(\ref{eqn:lagrange-dynamics1}). Assume that the initial conditions of the two systems  are the same. Denote the solutions of (\ref{eqn:lagrange-kinematics})-(\ref{eqn:lagrange-dynamics}) and (\ref{eqn:lagrange-kinematics1})-(\ref{eqn:lagrange-dynamics1}) as $(q(t),v(t))$ and $(\bar{q}(t),\bar{v}(t))$ for $t\ge 0$, respectively.
If the system (\ref{eqn:lagrange-kinematics1})-(\ref{eqn:lagrange-dynamics1}) is GAS with respect to $(\bar{x},\bar{y},\bar{v}_x,\bar{v}_y)$ uniformly in $(\bar{\theta}(0),\bar{\omega}(0))$, then the system (\ref{eqn:lagrange-kinematics})-(\ref{eqn:lagrange-dynamics}) is SPAS with respect to $({x},{y},{v}_x,{v}_y)$ uniformly in $({\theta}(0),{\omega}(0))$.
\end{theorem}
\begin{proof}
By the variation of constants formula in Appendix \ref{appendix:geometric}, the corresponding pull back system of (\ref{eqn:closed-loop}) is given by
\begin{equation}\label{eqn:pull-back2}
 \frac{{\rm d}}{{\rm d}\tau}
    \begin{bmatrix}
    \hat{q}\\\hat{v}
    \end{bmatrix}=\varepsilon \mathsf{F}(\tau,\hat{q},\hat{v}),
\end{equation}
where $(\hat{q}(0),\hat{v}(0))=(q(0),v(0))$ and
\begin{equation*}
\begin{split}
&\mathsf{F}(\tau,q,v)=\mathsf{f}(q,v)\\
&\, +\sum_{k=1}^\infty \int_0^\tau\cdots\int_0^{s_{k-1}}\left({\rm ad}_{\mathsf{g}(s_k,q)}\cdots{\rm ad}_{\mathsf{g}(s_1,q)}\mathsf{f}(q,v)\right) {\rm d}s_k\cdots{\rm d}s_1.
\end{split}
\end{equation*}
By direct calculations, we have
\begin{equation*}
\begin{split}
    &{\rm ad}_{\mathsf{g}(s_1,z)}\mathsf{f}(q,v)=\sum_{i=1}^m w_i(s_1)
    \begin{bmatrix}
    J(q)B_i(q)\\
    \left(\dfrac{\partial \mathsf{f}_2}{\partial v}\right) B_i-\left(\dfrac{\partial B_i}{\partial q}\right)J(q)v
    \end{bmatrix},
\end{split}
\end{equation*}
and
\begin{equation*}
{\rm ad}_{\mathsf{g}(s_2,q)}{\rm ad}_{\mathsf{g}(s_1,q)}\mathsf{f}(q,v)=-\sum_{i,j=1}^m w_i(s_1)w_j(s_2)
    \begin{bmatrix}
    0\\
    \langle B_i:B_j \rangle
    \end{bmatrix}.
\end{equation*}
Note that the symmetric product $\langle B_i:B_j \rangle$ is a vector field depending only on $q$. Thus, the higher order terms ${\rm ad}_{\mathsf{g}(s_k,q)}\cdots{\rm ad}_{\mathsf{g}(s_1,q)}\mathsf{f}(q,v)\equiv 0$ for all $k\ge 3$. The pull back vector field $\mathsf{F}$ is given by
\begin{align}
\mathsf{F}&=\mathsf{f}+
\sum_{i=1}^m \begin{bmatrix}
    J(q)B_i(q)\\
    \left(\dfrac{\partial \mathsf{f}_2}{\partial v}\right) B_i-\left(\dfrac{\partial B_i}{\partial q}\right)J(q)v
    \end{bmatrix}
    \int_{0}^{\tau}w_i(s_1){\rm d}s_1 \notag\\
&\quad -\sum_{i,j=1}^m
    \begin{bmatrix}
    0\\
    \langle B_i:B_j \rangle
    \end{bmatrix}\int_{0}^\tau \int_{0}^{s_1} w_i(s_1)w_j(s_2){\rm d}s_2{\rm d}s_1.\label{eqn:pull-back3}
\end{align}
Note that the system (\ref{eqn:closed-loop}) is in the form of (\ref{eqn:voc}), and its pull back system is given by (\ref{eqn:pull-back2})-(\ref{eqn:pull-back3}). Denote the solution of the pull back system (\ref{eqn:pull-back2}) by $(\hat{q}(\tau),\hat{v}(\tau))$. Then, it follows from Theorem \ref{thm:variation-of-constants} that the solution of (\ref{eqn:closed-loop}) is given by the initial value problem
\begin{equation}
    \frac{{\rm d}}{{\rm d}\tau}
    \begin{bmatrix}
    q\\v
    \end{bmatrix}=
    \begin{bmatrix}
    0\\ \sum\limits_{i=1}^{m}B_i(q)w_i(\tau)
    \end{bmatrix},\quad
    \begin{bmatrix}
    q(0)\\v(0)
    \end{bmatrix}=\begin{bmatrix}
    \hat{q}(\tau)\\\hat{v}(\tau)
    \end{bmatrix}.
\end{equation}
Therefore, we have
\begin{align}
q(\tau)& = q(0)= \hat{q}(\tau), \label{eqn:soln1}\\
v(\tau)&=\hat{v}(\tau)+\Xi(\tau,q(\tau)). \label{eqn:soln2}
\end{align}
The pull back system (\ref{eqn:pull-back2}) is in the classical averaging form \cite[Section 10.4]{khalil2002nonlinear}. Consider the average system
\begin{equation}\label{eqn:averaged1}
 \frac{{\rm d}}{{\rm d}\tau}
    \begin{bmatrix}
    \bar{q}\\\bar{v}
    \end{bmatrix}=\frac{\varepsilon}{T}\int_{0}^T \mathsf{F}(\tau,\bar{q},\bar{v}) {\rm d}\tau,
\end{equation}
and denote the solution by $(\bar{q}(t),\bar{v}(t))$. It follows from (\ref{eqn:periodic-2}), the symmetry of the symmetric product, and integration by parts, that the averaged system (\ref{eqn:averaged1}) in time scale $t=\varepsilon\tau$ is the symmetric product system (\ref{eqn:lagrange-kinematics1})-(\ref{eqn:lagrange-dynamics1}). 

According to the averaging theorem \cite[Theorem 10.4]{khalil2002nonlinear}, there exists $\varepsilon^*>0$ such that for all $0<\varepsilon<\varepsilon^*$, 
\begin{equation}\label{eqn:approximation}
|\hat{q}(t)-\bar{q}(t)|=O(\varepsilon),\quad {\rm and}\quad |\hat{v}(t)-\bar{v}(t)|=O(\varepsilon)
\end{equation}
as $\varepsilon\to 0$ on time scale 1. We recover the partial converging trajectories property by substituting (\ref{eqn:soln1})-(\ref{eqn:soln2}) into (\ref{eqn:approximation}). Finally, it follows directly from Proposition \ref{thm:partial-SPUAS} that the system (\ref{eqn:lagrange-kinematics})-(\ref{eqn:lagrange-dynamics}) is SPAS with respect to $({x},{y},{v}_x,{v}_y)$ uniformly in $({\theta}(0),{\omega}(0))$, which completes the proof.
\end{proof}

\begin{remark}\rm
In Theorem \ref{thm:symmetric-product}, instead of requiring  UGAS of the symmetric product system (\ref{eqn:lagrange-kinematics1})-(\ref{eqn:lagrange-dynamics1}) as in classical averaging theory \cite{teel1999semi}, we only assume (\ref{eqn:lagrange-kinematics1})-(\ref{eqn:lagrange-dynamics1}) to be P-UGAS with respect to $(\bar{x},\bar{y},\bar{v}_x,\bar{v}_y)$, while the remaining part of the state $(\bar{\theta}(t),\bar{\omega}(t))$ does not necessarily converge to $(0,0)$. Correspondingly, in the source seeking design in the next section, the  approximation in the linear motion $|({x},{y},{v}_x,{v}_y)-(x^\star,y^\star,0,0)|=O(\varepsilon)$ is valid for all $t\ge 0$, while the angular motion of the vehicle can be sustained.
\end{remark}

\begin{remark}\rm
The total energy of the planar vehicle is $E=\frac{1}{2}v^\top Mv$.
If the vector fields $B_i(q)$ are integrable, they can be written as $B_i(q)=\nabla \varphi_i(q)$ for some scalar functions $\varphi_i(q)$, where $i=1,\ldots,m$. It follows from \cite{crouch1981geometric} that if we define \textit{symmetric product for two scalar functions} (Beltrami bracket) according to $\langle \varphi_i:\varphi_j \rangle\coloneqq (\nabla \varphi_i)^\top\nabla \varphi_j$, then we have $\nabla\langle \varphi_i:\varphi_j \rangle(q)=\langle \nabla\varphi_i:\nabla\varphi_j \rangle(q)=\langle B_i:B_j \rangle(q)$. Correspondingly, the total energy of the symmetric product system (\ref{eqn:lagrange-kinematics1})-(\ref{eqn:lagrange-dynamics1}) is $E_{\mathtt{av}}=\frac{1}{2}\bar{v}^\top M\bar{v} + \sum_{i,j=1}^m \Lambda_{ij}\langle \varphi_i:\varphi_j \rangle(\bar{q})$. The term $\sum_{i,j=1}^m \Lambda_{ij}\langle \varphi_i:\varphi_j \rangle(\bar{q})$, introduced by the ``high-magnitude high-frequency forces", is called the \textit{averaged potential} \cite{bullo2000controllability}. Thus, the control law (\ref{eqn:input}) can be viewed as a ``potential energy shaping" technique, where the desired potential energy function can be injected by designing appropriate input vectors $B_i(q)$. This viewpoint shows that besides the classical averaging approach in \cite{krstic2000stability} and the Lie bracket averaging approach in \cite{durr2013lie}, the symmetric product approximation can also be used to obtain gradient information, which will be used in the source seeking design.
\end{remark}

\section{Source Seeking for Underactuated Vehicles}\label{sec:controldesign}

\subsection{Source Seeking Scheme}
\begin{figure}[t]
	\centering
	\includegraphics[scale=0.4]{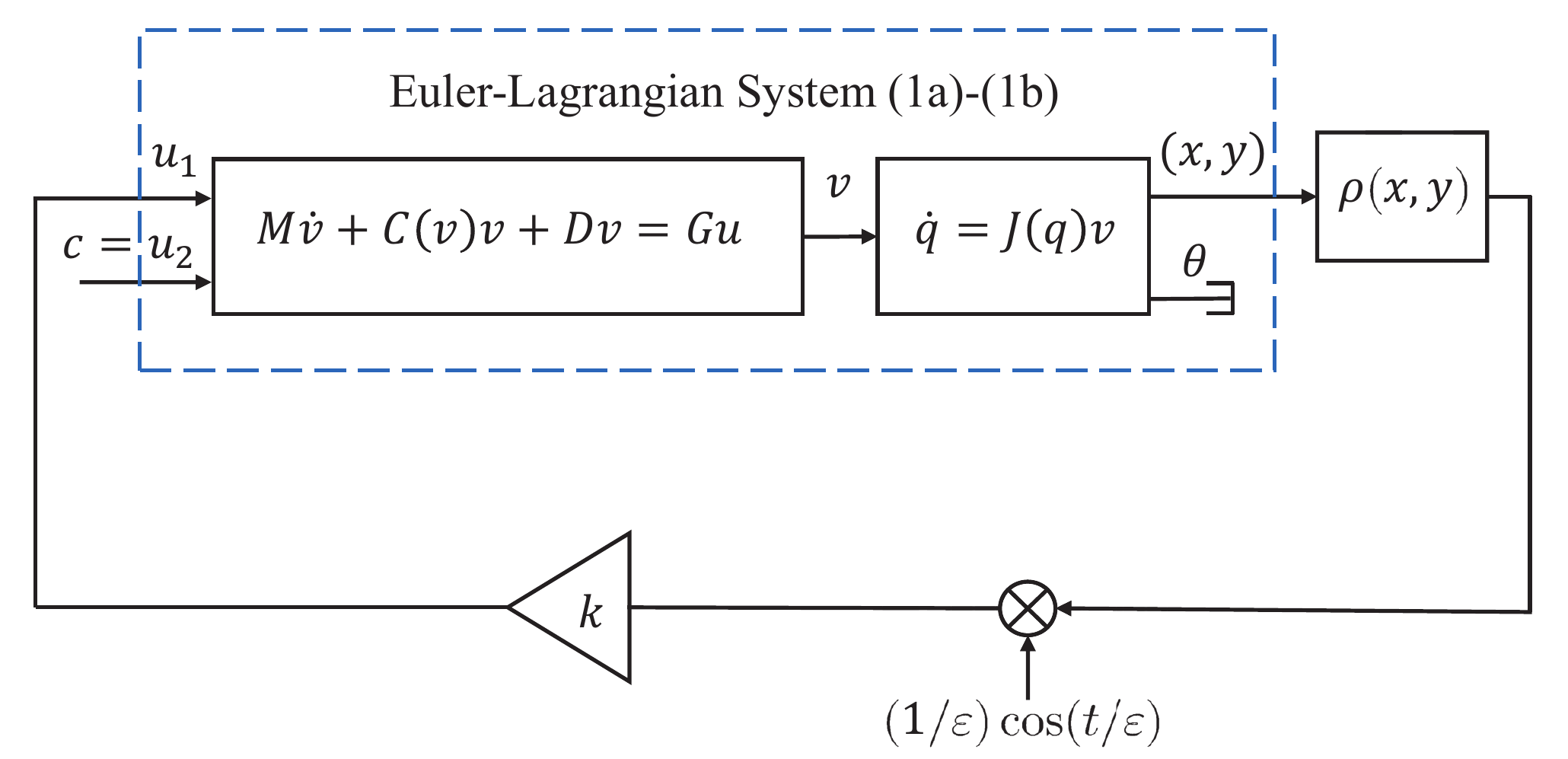}
	\caption{Source seeking scheme for planar vehicle system (\ref{eqn:lagrange-kinematics})-(\ref{eqn:lagrange-dynamics}).}
	\label{fig:ESC}
\end{figure}

We propose a source seeking scheme for the planar vehicle system (\ref{eqn:lagrange-kinematics})-(\ref{eqn:lagrange-dynamics}) as it is depicted in Fig. \ref{fig:ESC}. In the proposed scheme, the surge force of the vehicle is tuned by the ES loop, while the yaw torque keeps a certain constant value. The proposed surge force tuning based source seeking scheme is similar to the methods in \cite{zhang2007source,durr2013lie,suttner2020acceleration}, but will be analyzed in the symmetric product approximation framework.

The control law in Fig. \ref{fig:ESC} is given by
\begin{align}
u_1&=\frac{k}{\varepsilon}\cos\left(\frac{t}{\varepsilon}\right)\rho(x,y), \label{eqn:input1}\\
u_2&=c,\label{eqn:input2}
\end{align}
where $\varepsilon$, $k$, and $c$ are positive parameters.  The gain $k$ is used to tune the transient performance. The small parameter $\varepsilon$ introduces the ``high-magnitude high-frequency force", which leads to the symmetric product approximation. The constant torque $c$ maintains a sustained angular motion of the vehicle, which is necessary to establish convergence for underactuated vehicle systems.

\subsection{Stability Analysis}

\begin{theorem}
Consider the system (\ref{eqn:lagrange-kinematics})-(\ref{eqn:lagrange-dynamics})  with inputs (\ref{eqn:input1})-(\ref{eqn:input2}). Suppose that the cost function $\rho(x,y)\ge 0$ satisfies (\ref{eqn:rho}). Then, for any $c\in(0,\hat{c})$, $\hat{c}>0$, and any $k>0$, there exists $\hat{\varepsilon}>0$ such that for the given $c,k$, and any $\varepsilon\in(0,\hat{\varepsilon})$, the closed-loop system is SPAS with respect to $({x}-x^\star,{y}-y^\star,{v}_x,{v}_y)$ uniformly in $({\theta}(0),{\omega}(0))$.
\end{theorem}
\begin{proof}
Note that the control law (\ref{eqn:input1})-(\ref{eqn:input2}) is in the form of (\ref{eqn:input}), where $m=1$, $b_0=[0,c]^\top$, $b_1(q)=[k\rho(x,y),0]^\top$, and $w_1(t)=\cos(t)$. It can be verified that conditions (\ref{eqn:periodic-1})-(\ref{eqn:periodic-2}) hold for $T=2\pi$. Thus, it follow from Theorem \ref{thm:symmetric-product}  that the closed-loop system is SPAS with respect to $({x}-x^\star,{y}-y^\star,{v}_x,{v}_y)$ uniformly in $({\theta}(0),{\omega}(0))$ if the corresponding symmetric product system (\ref{eqn:lagrange-kinematics1})-(\ref{eqn:lagrange-dynamics1}) is GAS with respect to $(\bar{x}-x^\star,\bar{y}-y^\star,\bar{v}_x,\bar{v}_y)$ uniformly in $({\theta}(0),{\omega}(0))$.  Next, we show that it is indeed the case.

By direct calculations, we have  $\Lambda_{11}=1/4$, and the symmetric product is given by $\langle B_1:B_1\rangle(\bar{q})=2(m_{11}^{-1}k)^2\rho(\bar{x},\bar{y})[\rho_x'(\bar{x},\bar{y})\cos(\bar{\theta})+\rho_y'(\bar{x},\bar{y})\sin(\bar{\theta}),0,0]^\top$, where $\rho_x'(x,y)\coloneqq \partial \rho(x,y)/\partial x$ and $\rho_y'(x,y)\coloneqq \partial \rho(x,y)/\partial y$.  The constant torque $c$ is selected such that Proposition \ref{prop:shifted-passivity} holds, and then, the system (\ref{eqn:lagrange-kinematics1})-(\ref{eqn:lagrange-dynamics1}) is shifted passive under the steady-state input $u^*=b_0$. Denote $\alpha=(m^{-1}_{11}k)^2/2$ and $C_i(\cdot)$ the $i$-th component of the vector $-M^{-1}C(\bar{v})\bar{v}$, and note that $M^{-1}D\bar{v}=[\frac{d_{11}}{m_{11}}\bar{v}_x,\frac{d_{22}}{m_{22}}\bar{v}_y,\frac{d_{33}}{m_{33}}\bar{\omega}]^\top$, $M^{-1}B_{0}=[0,0,\frac{c}{m_{33}}]^\top$.
The symmetric product system can be viewed as a feedback interconnection of two subsystems, as shown in Fig. \ref{fig:sketch}.

When the input $\bar{v}_y\equiv 0$, the nominal system of the upper subsystem is exactly the unicycle model under passive feedback. We first prove the nominal system of the upper subsystem (i.e., $\bar{v}_y\equiv 0$) is P-UGAS. Let $V_1(\bar{x},\bar{y},\bar{v}_x)=\frac{1}{2}\bar{v}_x^2 + \frac{\alpha}{2}\rho(\bar{x},\bar{y})^2$, and along trajectories of the nominal system, we have $\dot{V}_1 |_{\tt nominal}=-\frac{d_{11}}{m_{11}}\bar{v}_x^2\le 0$,
which, according to Theorem \ref{thm:P-UGAS}, shows that the nominal system is US and UGB with respect to $(\bar{x}-x^\star,\bar{y}-y^\star,\bar{v}_x)$ uniformly in $(\bar{\theta}(0),\bar{\omega}(0))$. Then, consider the auxiliary function $V_2=\bar{v}_x \rho(\bar{x},\bar{y}) (\rho_x'\cos(\bar{\theta})+\rho_y'\sin(\bar{\theta}))$. Evaluating the time derivative of $V_2$ along trajectories of the nominal system on the set $\{\bar{v}_x=0\}$, we have $\dot{V}_2 |_{{\tt nominal, } \bar{v}_x=0}=-\alpha\rho^2(\rho_x'\cos(\bar{\theta})+\rho_y'\sin(\bar{\theta}) )^2$, which is non-zero definite. It follows from Matrosov' theorem \cite{wang2021leader,hahn1967stability} that the nominal system is UGAS with respect to $(\bar{x}-x^\star,\bar{y}-y^\star,\bar{v}_x)$ uniformly in $(\bar{\theta}(0),\bar{\omega}(0))$.

Second, we prove that the upper subsystem is input-to-output stable (IOS) by viewing $\bar{v}_y$ as input and $(\bar{v}_x,\bar{\omega})$ as output. Because the nominal part of the upper subsystem is P-UGAS, for each $r>0$, there exists a constant $\delta_r>0$ such that for all initial conditions starting in the ball centering at the equilibrium with radius $r$, we have $\max\{|\rho_x'\cos(\bar{\theta})+\rho_y'\sin(\bar{\theta}) |,|\rho_x'\cos(\bar{\theta})+\rho_y'\sin(\bar{\theta}) |^2, |c\rho(\rho_x'\sin(\bar{\theta})-\rho_y'\cos(\bar{\theta}))|/d_{33}, |\rho(\rho_{xx}''\cos(\bar{\theta})^2+2\rho_{xy}''\sin(\bar{\theta})) \cos(\bar{\theta})|\}<\delta_r$. Let $\mathcal{V}_r=\beta_r V_1 + V_2$, where $\beta_r>0$ is a constant to be determined. It follows from Young's inequality
$ab\le a^2/(2\epsilon) + (\epsilon b^2)/2$ that $\mathcal{V}_r>0$  and $\dot{\mathcal{V}_r}|_{\tt nominal}\le -\bar{v}_x^2-\frac{\alpha}{2}\rho^2(\rho_x'\cos(\bar{\theta})+\rho_y'\sin(\bar{\theta}) )^2 + \bar{v}_x\delta_r$ by selecting $\beta_r>\max\left\{\delta^2_r/\alpha,1+2{m_{11}}\delta_r/{d_{11}} +d_{11}/(2\alpha m_{11})\right\}$. Then, taking time derivative of $\mathcal{V}_r$ along trajectories of the upper subsystem, and noting that the quadratic terms $-\bar{v}_x^2-\frac{\alpha}{2}\rho^2(\rho_x'\cos(\bar{\theta})+\rho_y'\sin(\bar{\theta}) )^2$ dominate $\dot{\mathcal{V}}_r|_{\tt upper}$ when $|(\bar{v}_x,\rho(\rho_x'\cos(\bar{\theta})+\rho_y'\sin(\bar{\theta}) ))|$ are large, we conclude that the upper subsystem is IOS with input $\bar{v}_y$ and output $(\bar{v}_x,\bar{\omega})$.

\begin{figure}[t]
	\centering
	\includegraphics[scale=0.5]{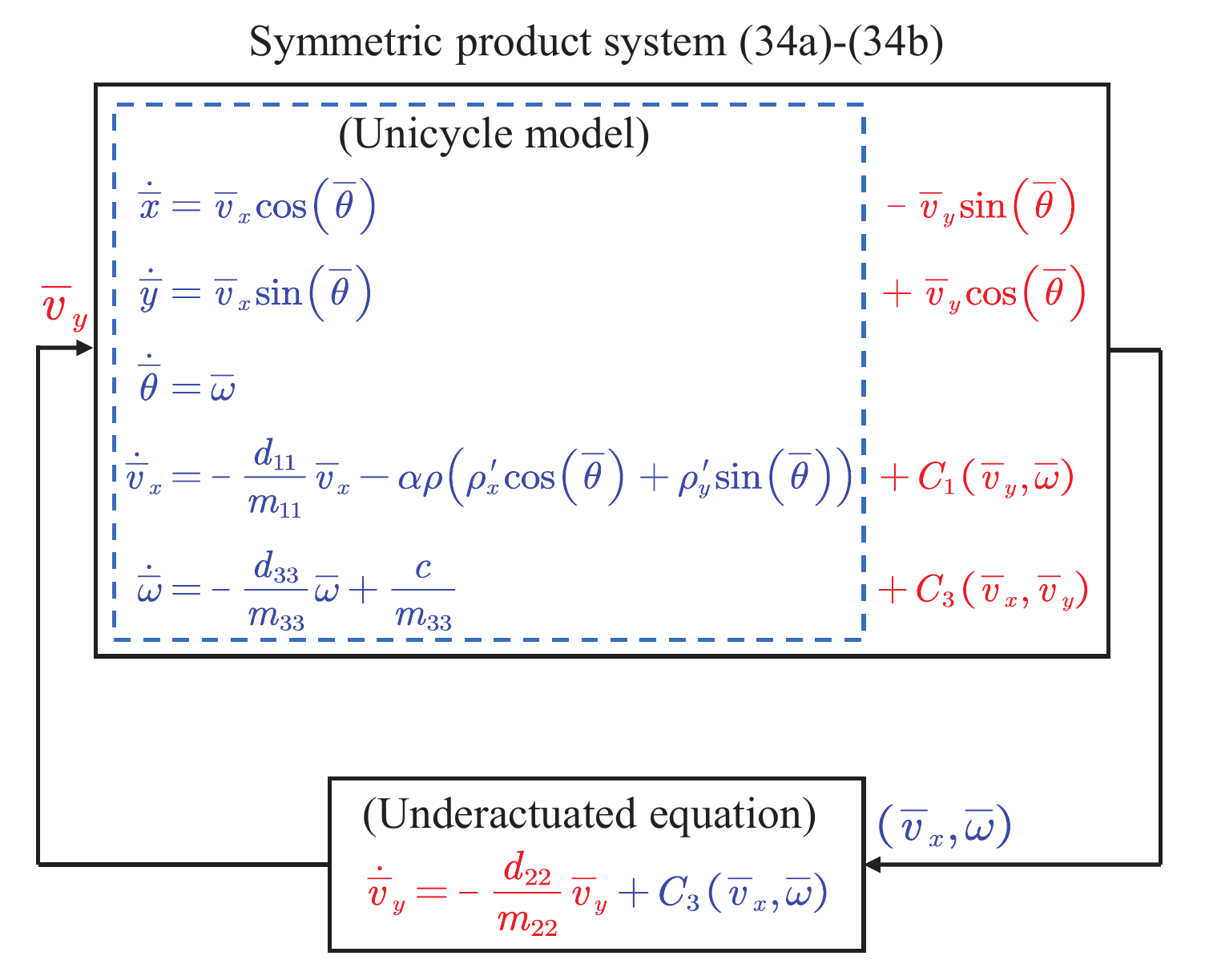}
	\caption{Feedback interconnection of the symmetric product system (\ref{eqn:lagrange-kinematics1})-(\ref{eqn:lagrange-dynamics1}).}
	\label{fig:sketch}
\end{figure}

Due to the fact that the lower subsystem in Fig. \ref{fig:sketch} is a stable linear system, it is also IOS by viewing $(\bar{v}_x,\bar{\omega})$ as the input and $\bar{v}_y$ as the output, and the IOS-gain can be rendered arbitrarily small by selecting $c$ small enough. Therefore, the symmetric product system (\ref{eqn:lagrange-kinematics1})-(\ref{eqn:lagrange-dynamics1}) is a feedback interconnection of two IOS subsystems, where the zero-state detectability can be easily verified. It follows from the small-gain theorem \cite{jiang1994small} that, there exists $\hat{c}>0$ such that the symmetric product system (\ref{eqn:lagrange-kinematics1})-(\ref{eqn:lagrange-dynamics1}) is GAS with respect to $(\bar{x}-x^\star,\bar{y}-y^\star,\bar{v}_x,\bar{v}_y)$ uniformly in $({\theta}(0),{\omega}(0))$ for all $c\in(0,\hat{c})$. Finally, we conclude that the  closed-loop system is SPAS with respect to $({x}-x^\star,{y}-y^\star,{v}_x,{v}_y)$ uniformly in $({\theta}(0),{\omega}(0))$ by invoking Theorem \ref{thm:symmetric-product}.
\end{proof}

\begin{remark}\rm 
Compared with the surge force tuning based source seeking schemes in  \cite{zhang2007source,durr2013lie}, the presented scheme does not require an additive periodic perturbation. The additive periodic perturbation is necessary for the Lie bracket averaging-based algorithm \cite{durr2013lie} since it is used to introduce the back-and-forth motion of a vehicle.
However, as shown in Section \ref{sec:approximations}, only with a multiplicative periodic perturbation, the pull back system still involves an operation that is calculating Lie bracket with the vector $\mathsf{g}$, i.e., $[\mathsf{g}(s_2,q),[\mathsf{g}(s_1,q),\mathsf{f}(q,v)]]$. We emphasize that the natural damping in the planar underactuated vehicle system (\ref{eqn:lagrange-kinematics})-(\ref{eqn:lagrange-dynamics}) plays an essential role in the source seeking design and stability analysis.
\end{remark}

\section{Simulations}
\label{sec:simulations}

Consider a boat with linear hydrodynamic damping \cite{wang2021leader}, where the equations are given by  (\ref{eqn:lagrange-kinematics})-(\ref{eqn:lagrange-dynamics}) with 
\begin{equation*}
C(v)=
\begin{bmatrix}
0& 0& -m_{22}v_y\\
0& 0& m_{11}v_x\\
m_{22}v_y & -m_{11}v_x & 0
\end{bmatrix}
\end{equation*}
and $D=\operatorname{diag}\{d_{11},d_{22},d_{33}\}$, where
\begin{align*}
&m_{11}=1.412,~~m_{22}=1.982,~~ m_{33}=0.354,\\
&d_{11}=3.436,~~~d_{22}=12.99,~~~ d_{33}=0.864.
\end{align*}
The boat is assumed to rest at the origin initially, i.e., $(q(0),v(0))=(0,0)$.
Assume that the cost function is $\rho(x,y)=(x - 2)^2 + 0.5(y - 3)^2 + 1$.

It follows from (\ref{eqn:lagrange-kinematics})-(\ref{eqn:lagrange-dynamics}) that the constant input $(u_1^*, u_2^*)=(0,c)$ leads to the steady-state velocity $v^*=(v_x^*,v_y^*,\omega^*)=(0,0,c/d_{33})$. Then, the constant $c$ is chosen such that $\partial \left[C(v)v^* \right]/\partial v + \left[\partial \left[C(v)v^* \right]/\partial v\right]^\top \le 2D$ holds. By direct calculation, we have $4d_{11}d_{22}-(\omega^*)^2(m_{11}-m_{22})^2\ge 0$, which implies that $c\le 2\sqrt{d_{11}d_{22}}d_{33}/(m_{22}-m_{11})=20.25$. That is, with the steady-state input $u^*=[0,c]^\top$ with $c\le 20.25$, the system is shifted passive.

\begin{figure}[!t]
     \centering
     \begin{subfigure}[b]{0.48\textwidth}
         \centering
         \includegraphics[scale=0.38]{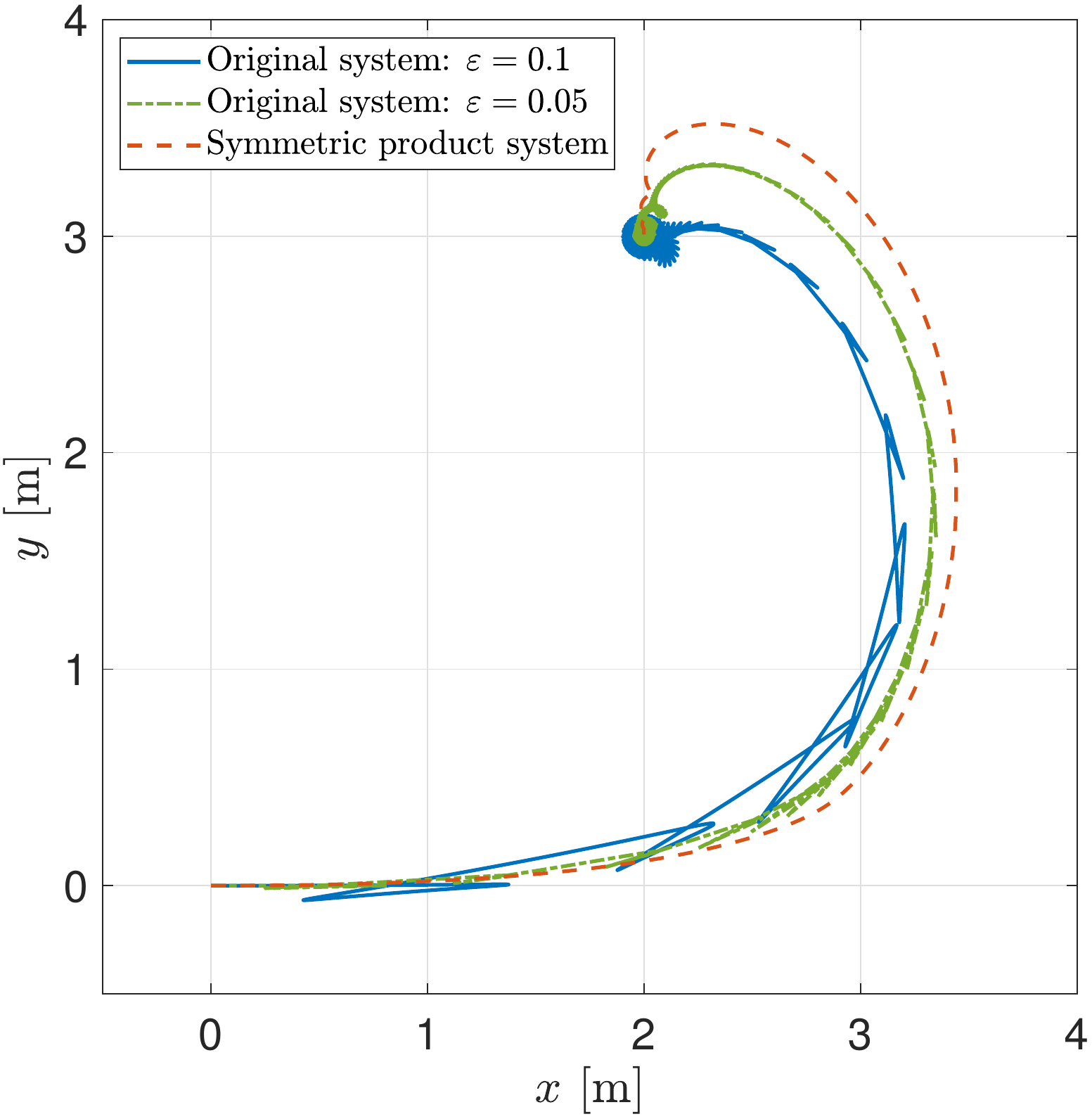}
         \caption{}
         \label{fig:path1}
     \end{subfigure}
     \hfill
     \begin{subfigure}[b]{0.48\textwidth}
         \centering
         \includegraphics[scale=0.35]{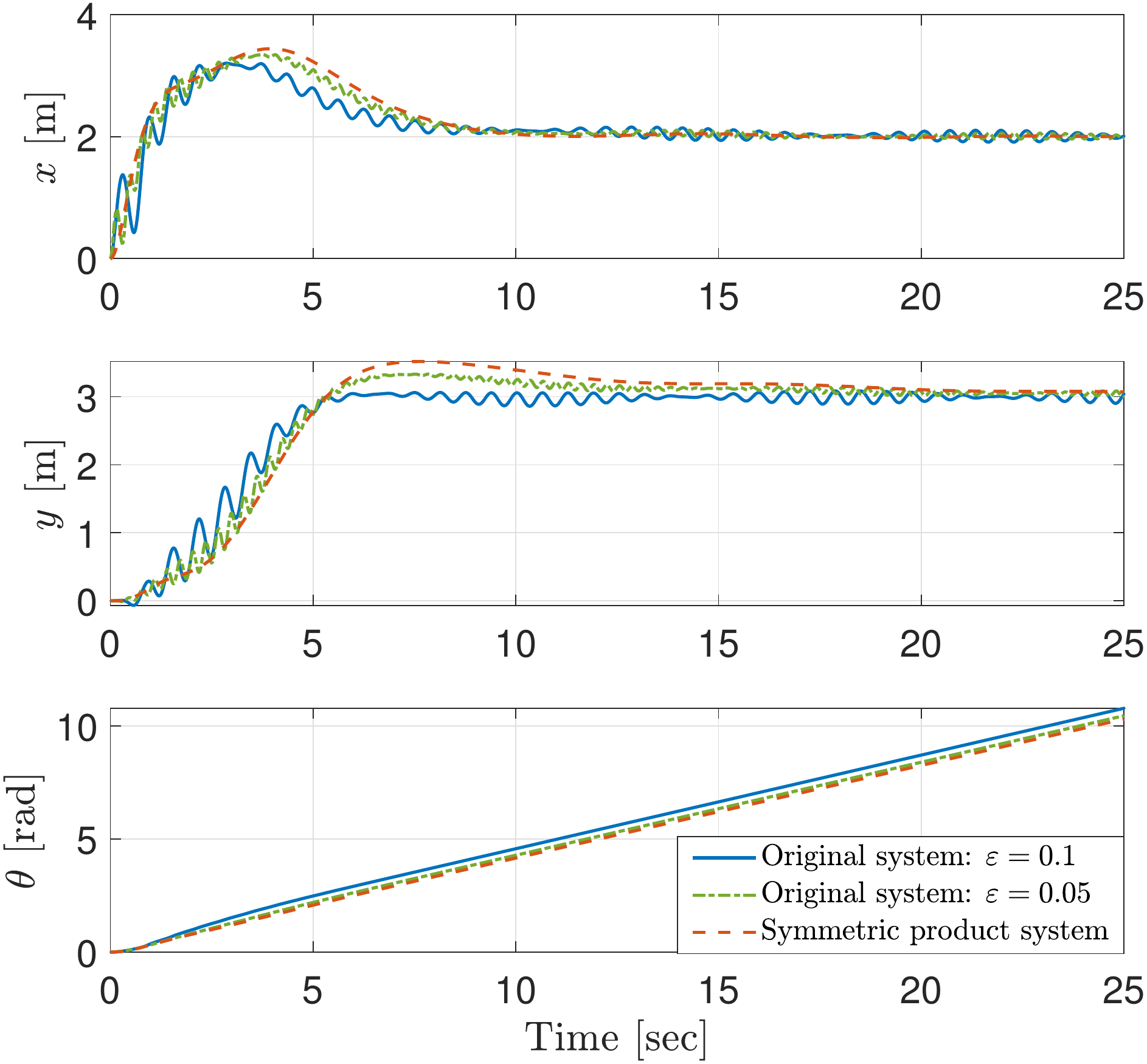}
         \caption{}
         \label{fig:trajectory1}
     \end{subfigure}
        \caption{Paths and configuration trajectories of the underactuated boat in source seeking $(c=1,k=1)$.}
        \label{fig:sim1}
\end{figure}

\begin{figure}[!t]
     \centering
     \begin{subfigure}[b]{0.48\textwidth}
         \centering
         \includegraphics[scale=0.38]{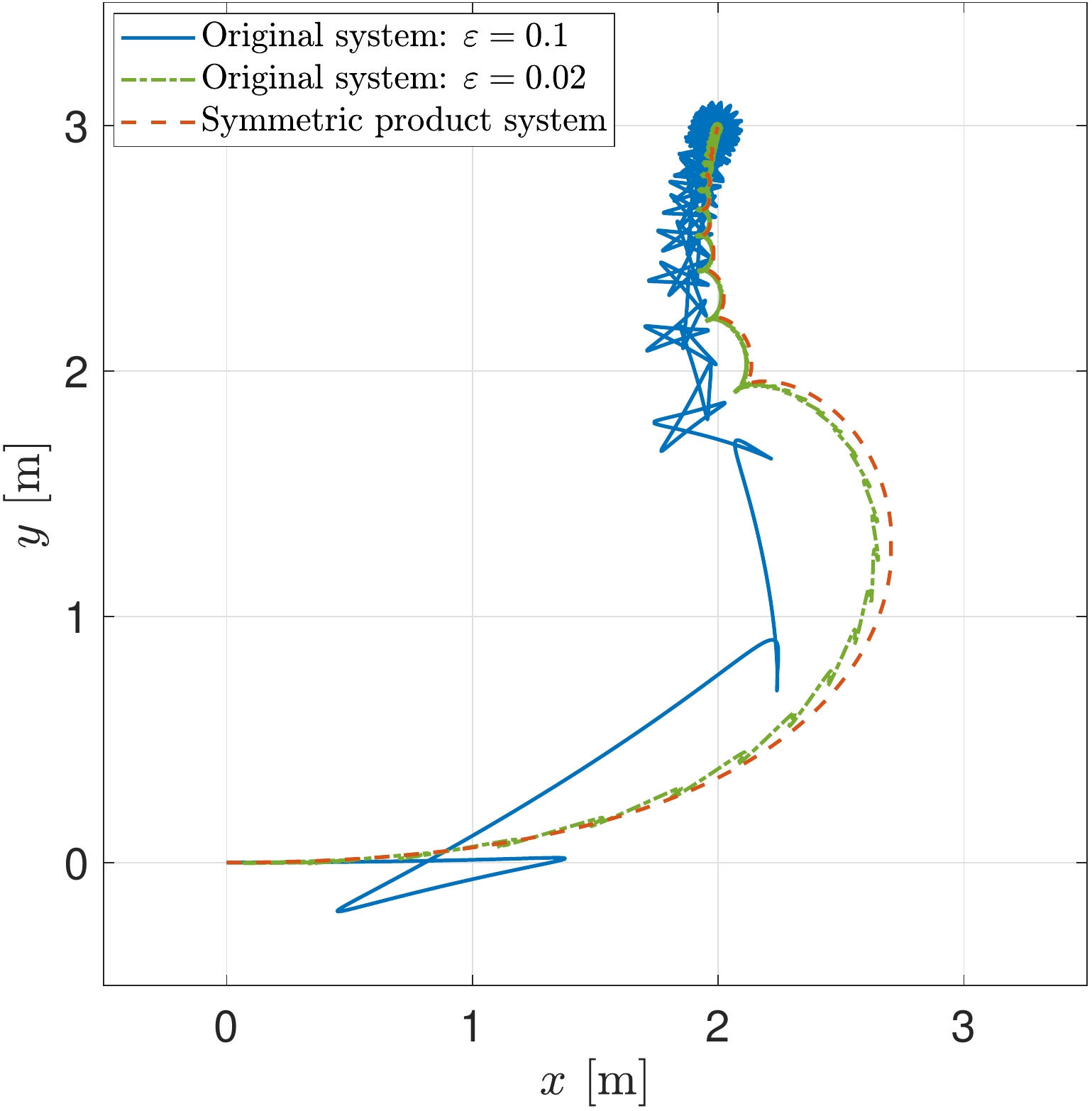}
         \caption{}
         \label{fig:path2}
     \end{subfigure}
     \hfill
     \begin{subfigure}[b]{0.48\textwidth}
         \centering
         \includegraphics[scale=0.35]{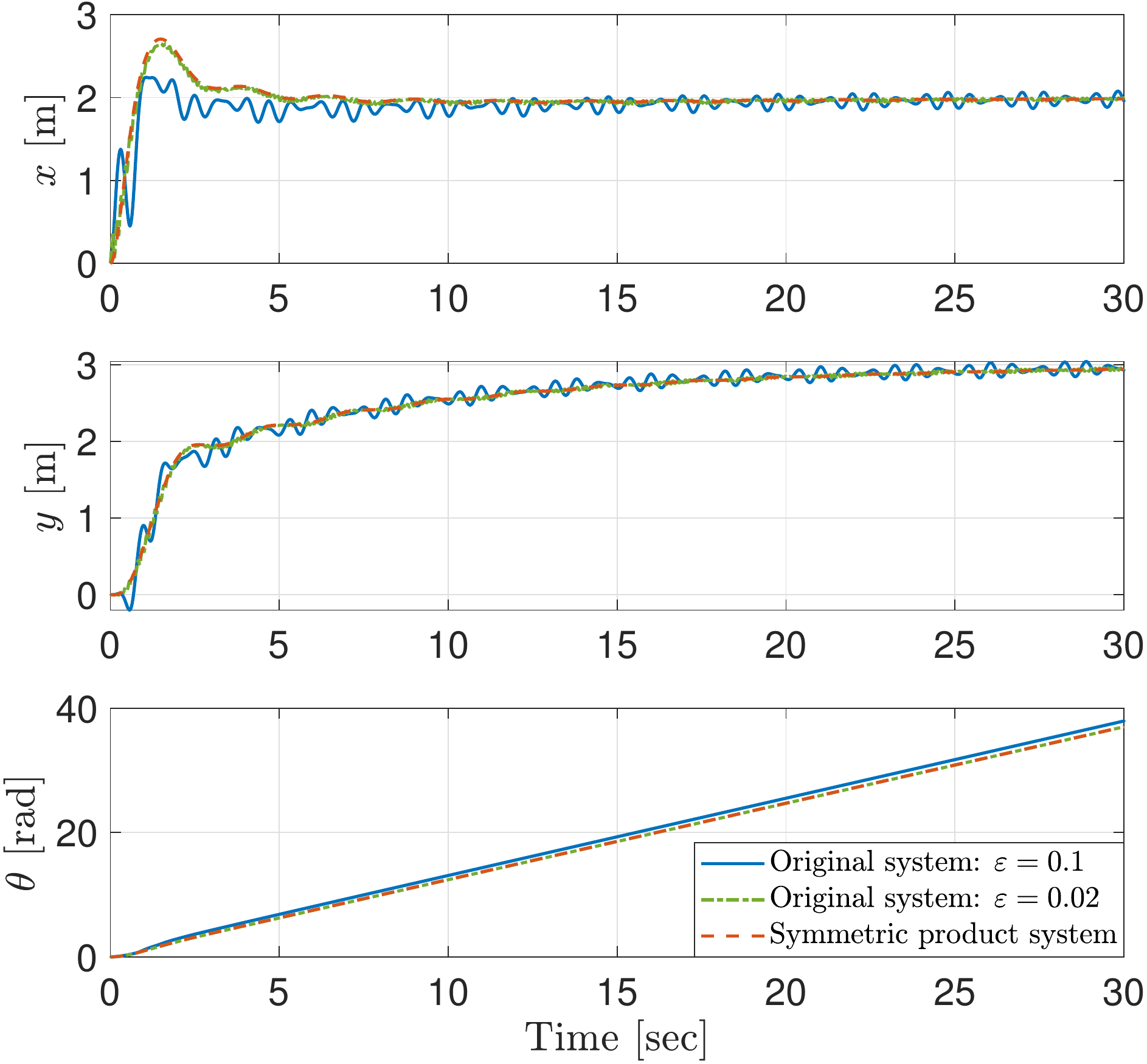}
         \caption{}
         \label{fig:trajectory2}
     \end{subfigure}
        \caption{Paths and configuration trajectories of the underactuated boat in source seeking $(c=3,k=1)$.}
        \label{fig:sim2}
\end{figure}

In the first example, we select the control parameters in (\ref{eqn:input1})-(\ref{eqn:input2}) to be $c=1$, $k=1$. The simulation results are shown in Fig. \ref{fig:sim1} for $\varepsilon=0.1$ and $\varepsilon=0.05$. In the second example, we increase the constant torque to $c=3$. The simulation results of the second example are shown in Fig. \ref{fig:sim2} for $\varepsilon=0.1$ and $\varepsilon=0.02$. It can be seen from both examples that the position trajectory of the underactuated boat converges to the $O(\varepsilon)$-neighborhood of the desired position $(x^\star,y^\star)=(2,3)$. Furthermore, as $\varepsilon\to 0$, the trajectories of the boat converge to the trajectory of the symmetric product system which represents the ideal solution. In general, a smaller $\varepsilon$ leads to a smoother trajectory. The only limitation on the value of $\varepsilon$ is the value of the control input (\ref{eqn:input1}) which increases as $\varepsilon$ decreases. In the third example, we select the control parameters to be $c=1$, $\varepsilon=0.1$. The simulation results are shown in Fig. \ref{fig:sim1} for $k=0.5$, $k=1$, and $k=1.5$. It can be seen from Fig.5 that increasing the gain $k$ may improve convergence speed but results in longer paths with more dramatic winding motion.

\begin{figure}[!t]
     \centering
     \begin{subfigure}[b]{0.48\textwidth}
         \centering
         \includegraphics[scale=0.33]{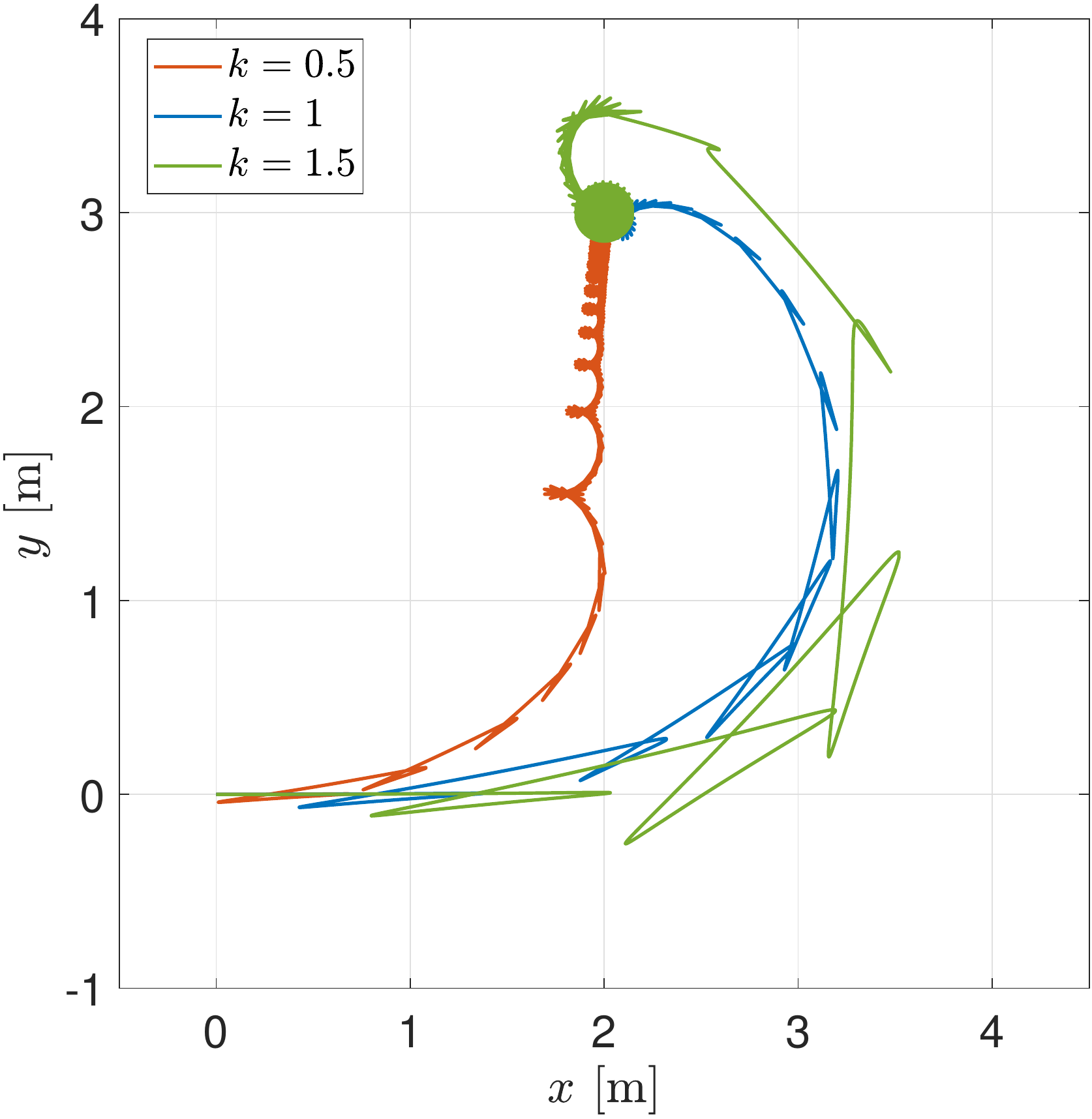}
         \caption{}
         \label{fig:path3}
     \end{subfigure}
     \hfill
     \begin{subfigure}[b]{0.48\textwidth}
         \centering
         \includegraphics[scale=0.34]{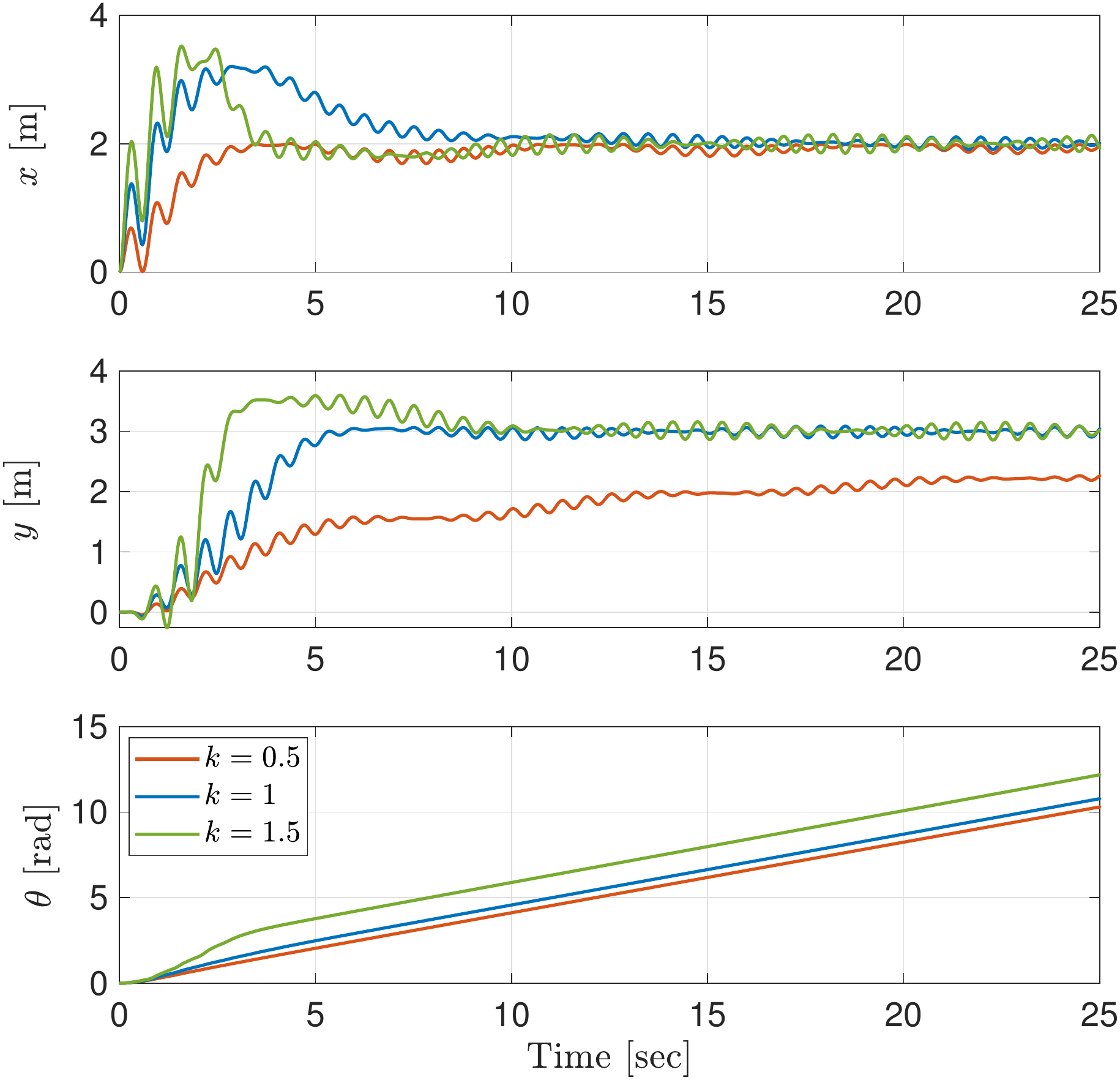}
         \caption{}
         \label{fig:trajectory3}
     \end{subfigure}
        \caption{Paths and configuration trajectories of the underactuated boat in source seeking $(c=1,\varepsilon=0.1)$.}
        \label{fig:sim3}
\end{figure}

\section{Experimental Results}
\label{sec:experiments}

\begin{figure}[!t]
     \centering
     \begin{subfigure}[b]{0.48\textwidth}
         \centering
         \includegraphics[scale=0.38]{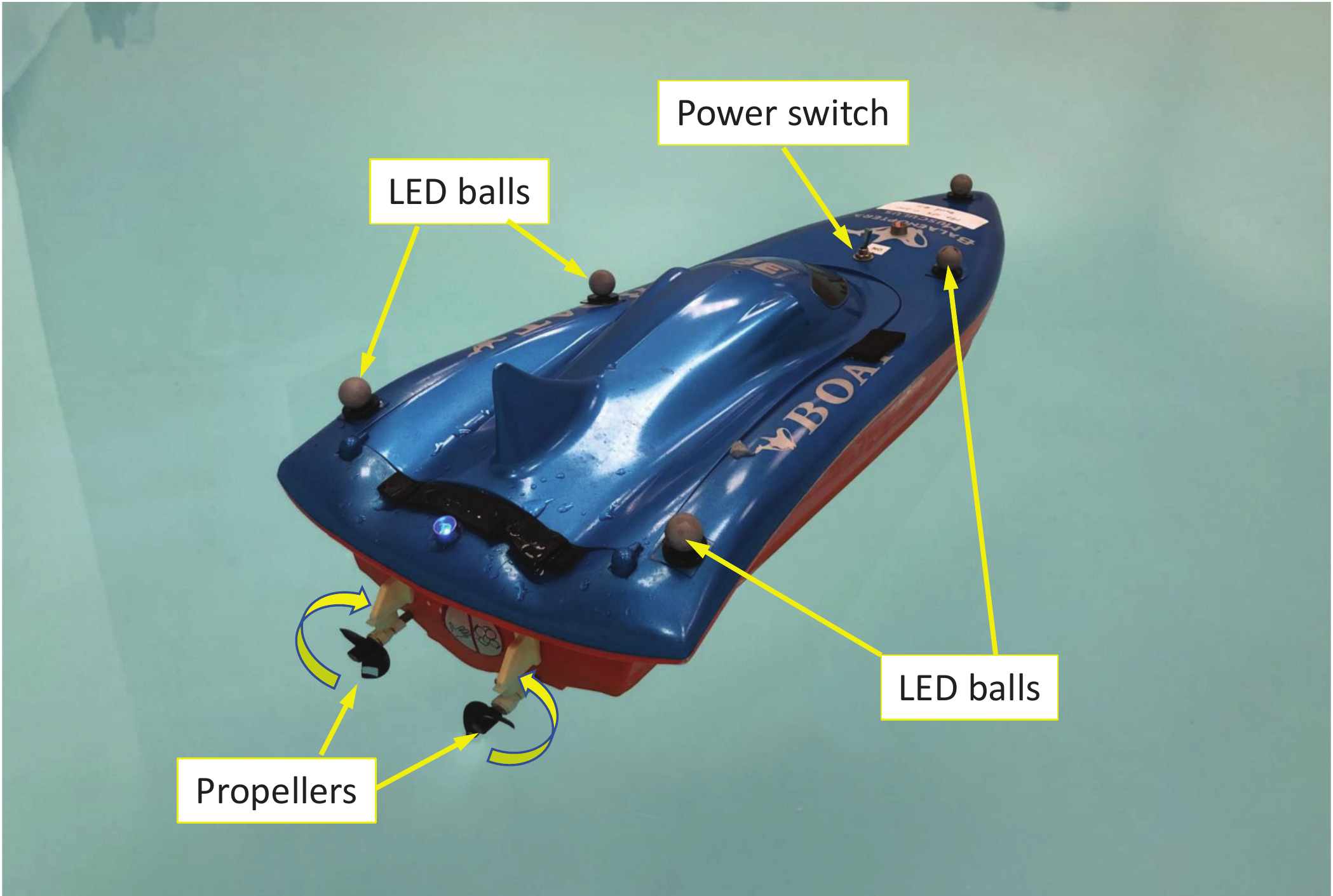}
         \caption{}
         \label{fig:setup1}
     \end{subfigure}
     \hfill
     \begin{subfigure}[b]{0.48\textwidth}
         \centering
         \includegraphics[scale=0.32]{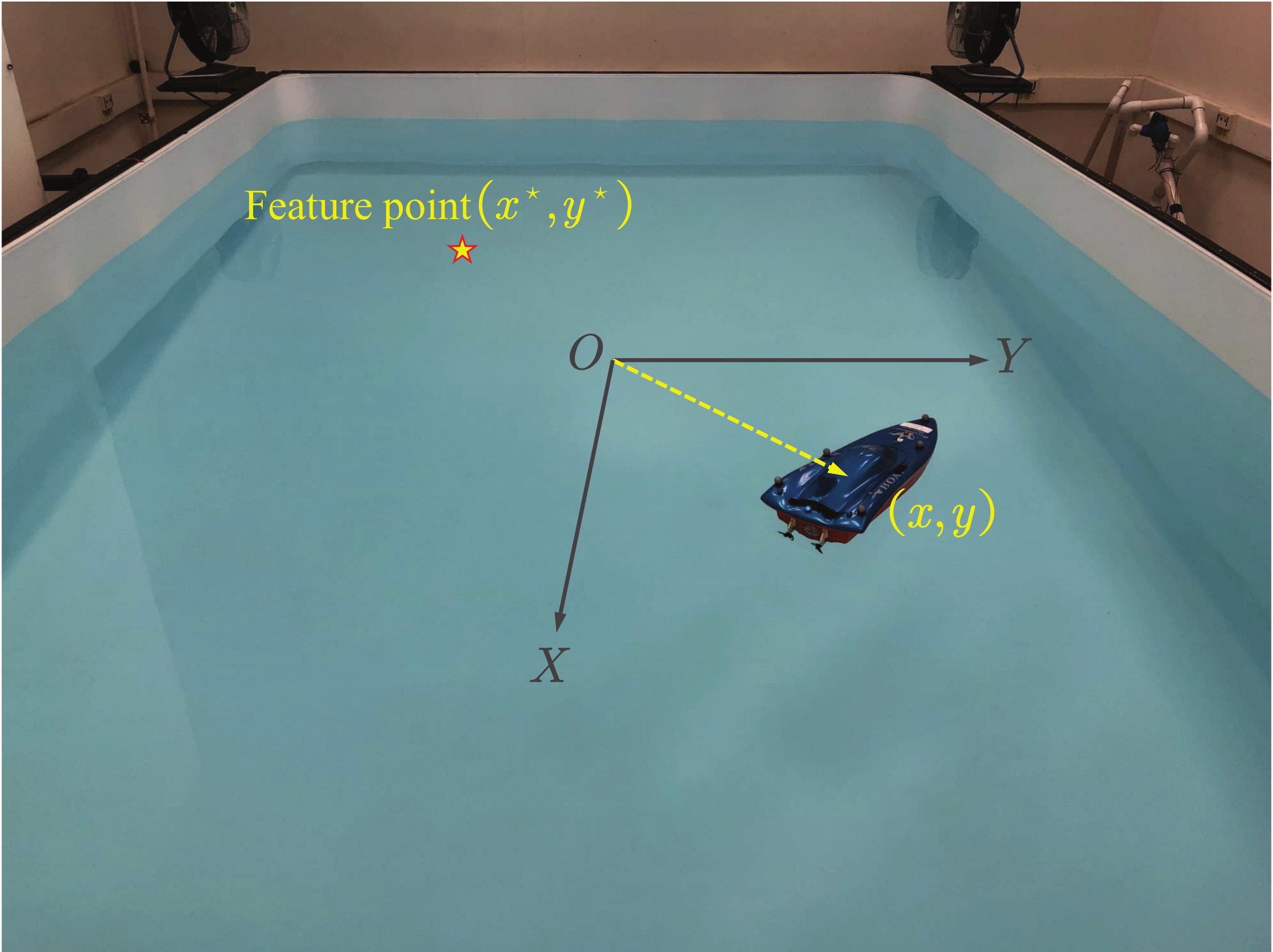}
         \caption{}
         \label{fig:setup2}
     \end{subfigure}
        \caption{The experimental boat in the water tank.}
        \label{fig:setup}
\end{figure}

To illustrate the algorithm's practicality and performance in the real world, experiments were performed with a small boat in the laboratory environment. The boat components include electronic speed control motors and propellers, a Raspberry Pi 3, a PWM driver, and a 3000 mAh lithium polymer battery. The boat operates in a pool equipped with the Vicon motion capture system, as shown in Fig. \ref{fig:setup}. The Vicon camera system captures infrared LED balls located on the boat and provides a relative distance between the feature point (source) and the boat to the control software implemented in MATLAB/Simulink. The control signals are sent to the Raspberry Pi via a Wi-Fi router.

\begin{figure}[ht]
     \centering
     \begin{subfigure}[b]{0.48\textwidth}
         \centering
         \includegraphics[scale=0.325]{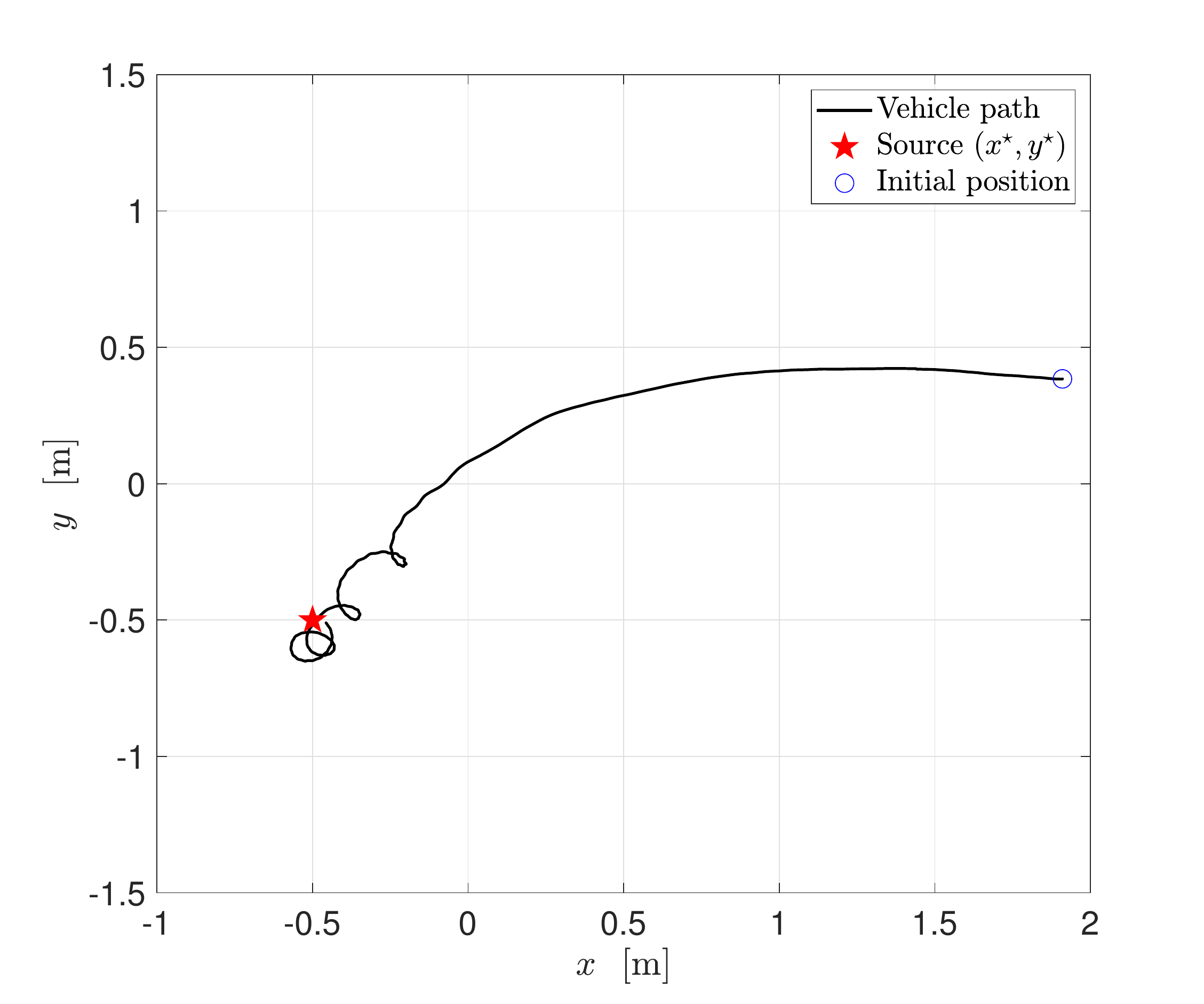}
         \caption{}
         \label{fig:path-1e}
     \end{subfigure}
     \hfill
     \begin{subfigure}[b]{0.48\textwidth}
         \centering
         \includegraphics[scale=0.325]{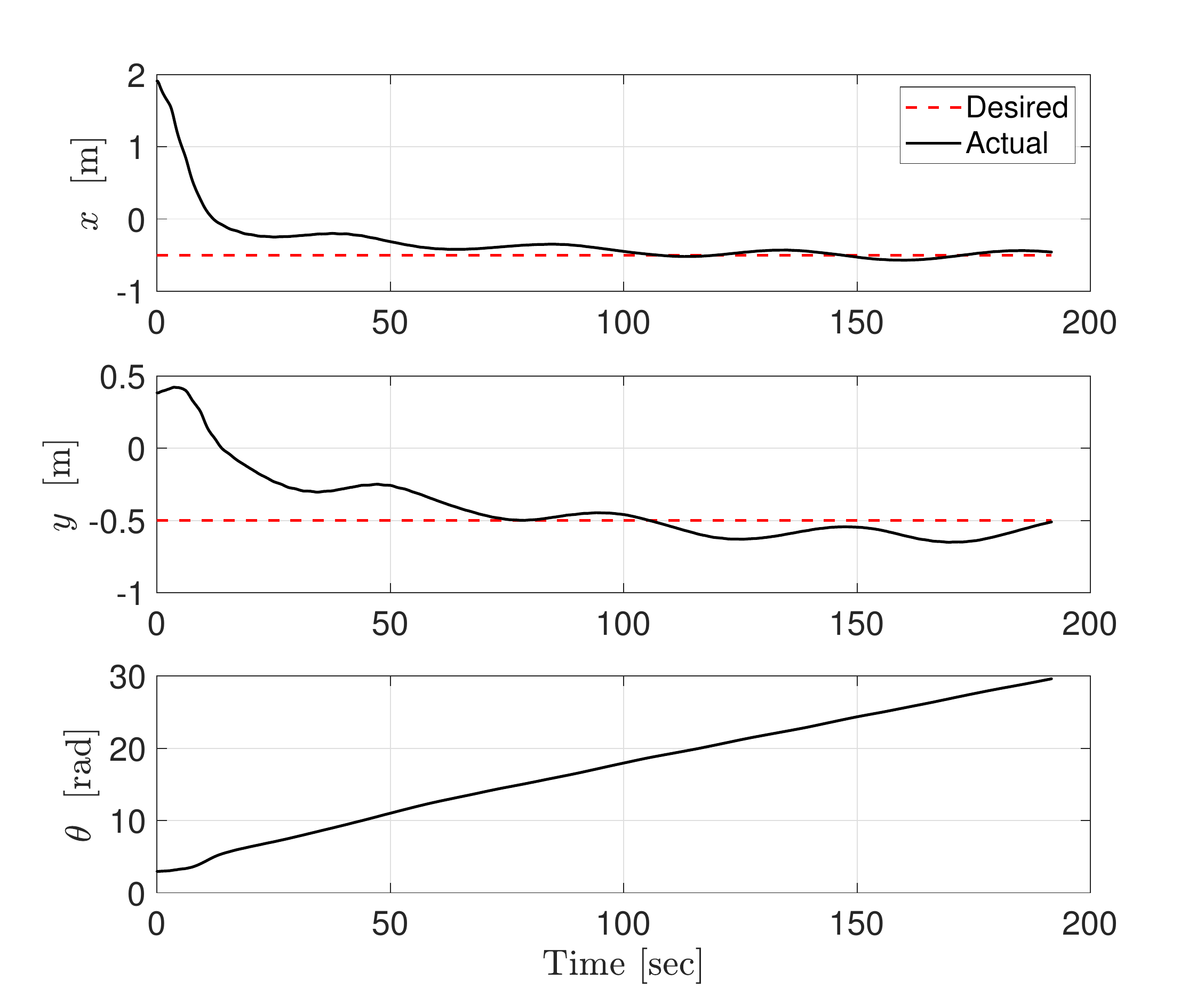}
         \caption{}
         \label{fig:traj-1e}
     \end{subfigure}
        \caption{Experimental path and configuration trajectories of the underactuated boat in source seeking (Case 1: $c=2.2\times 10^{-3},\varepsilon=0.5,k=0.08$).}
        \label{fig:experiments1}
\end{figure}

\begin{figure}[!t]
	\centering
	\includegraphics[scale=0.325]{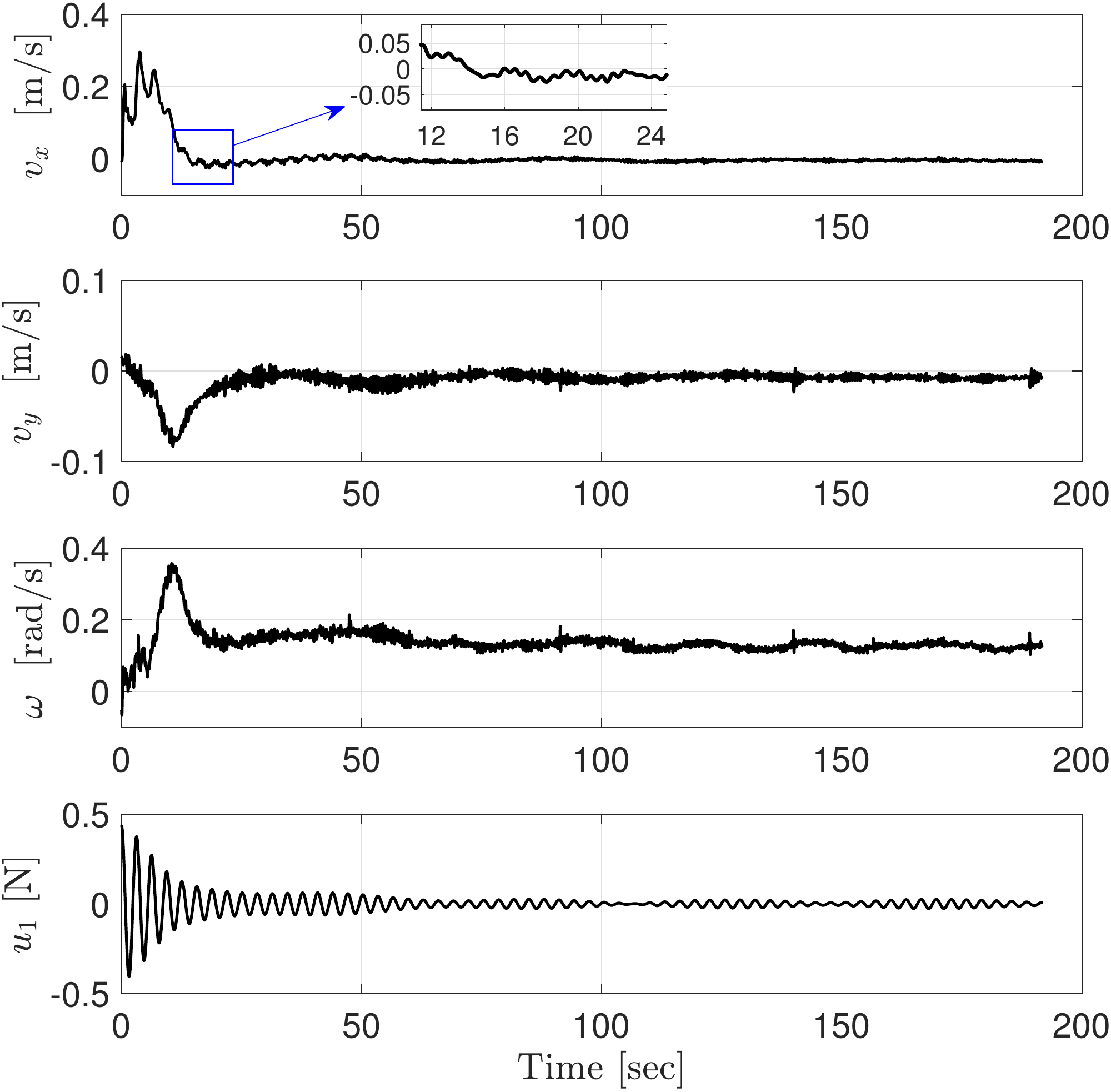}
	\caption{Experimental velocity trajectories and the surge force of the underactuated boat in source seeking (Case 1: $c=2.2\times10^{-3},\varepsilon=0.5,k=0.08$).}
	\label{fig:9d}
\end{figure}

\begin{figure}[!t]
     \centering
     \begin{subfigure}[b]{0.48\textwidth}
         \centering
         \includegraphics[scale=0.325]{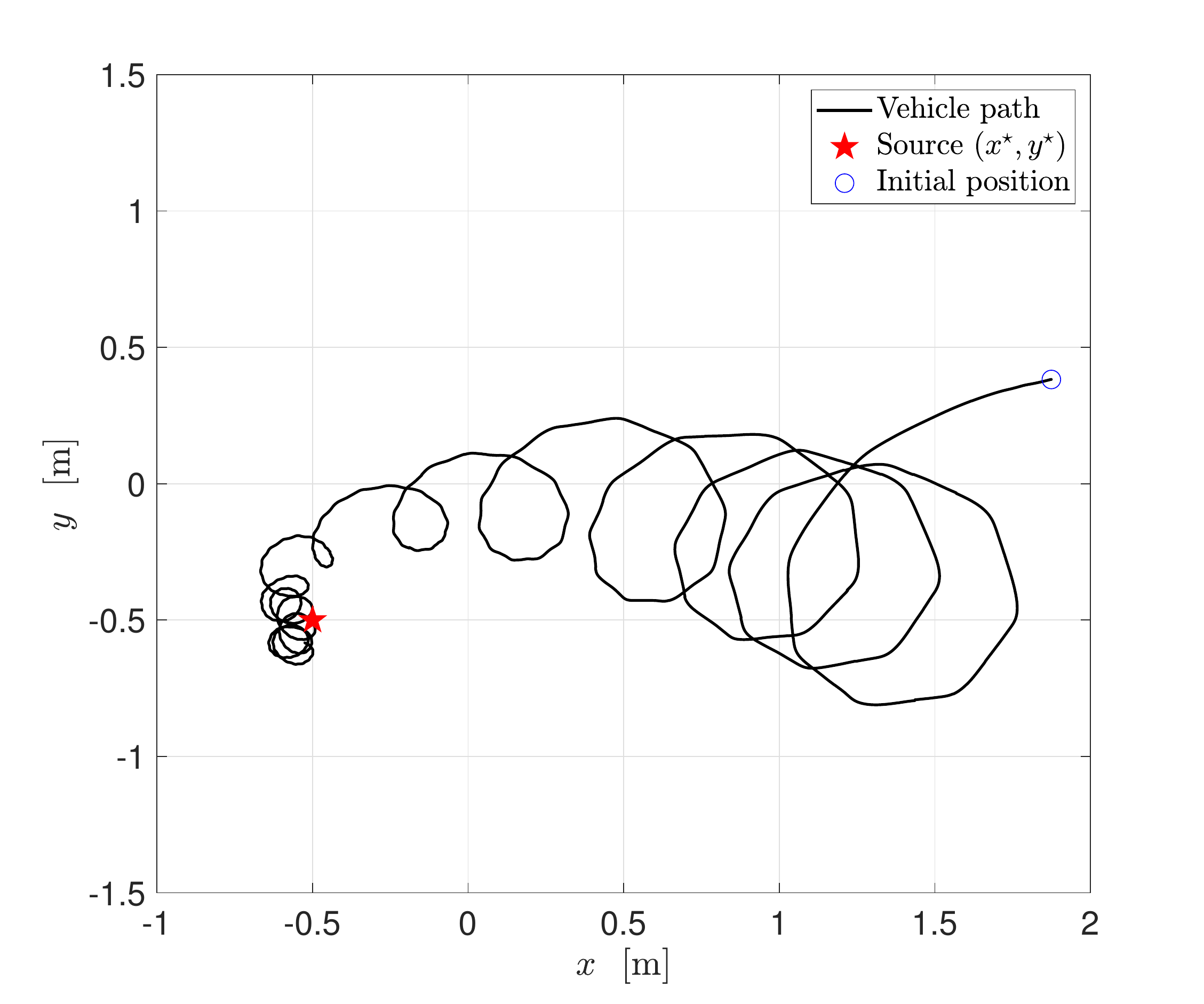}
         \caption{}
         \label{fig:path-2e}
     \end{subfigure}
     \hfill
     \begin{subfigure}[b]{0.48\textwidth}
         \centering
         \includegraphics[scale=0.325]{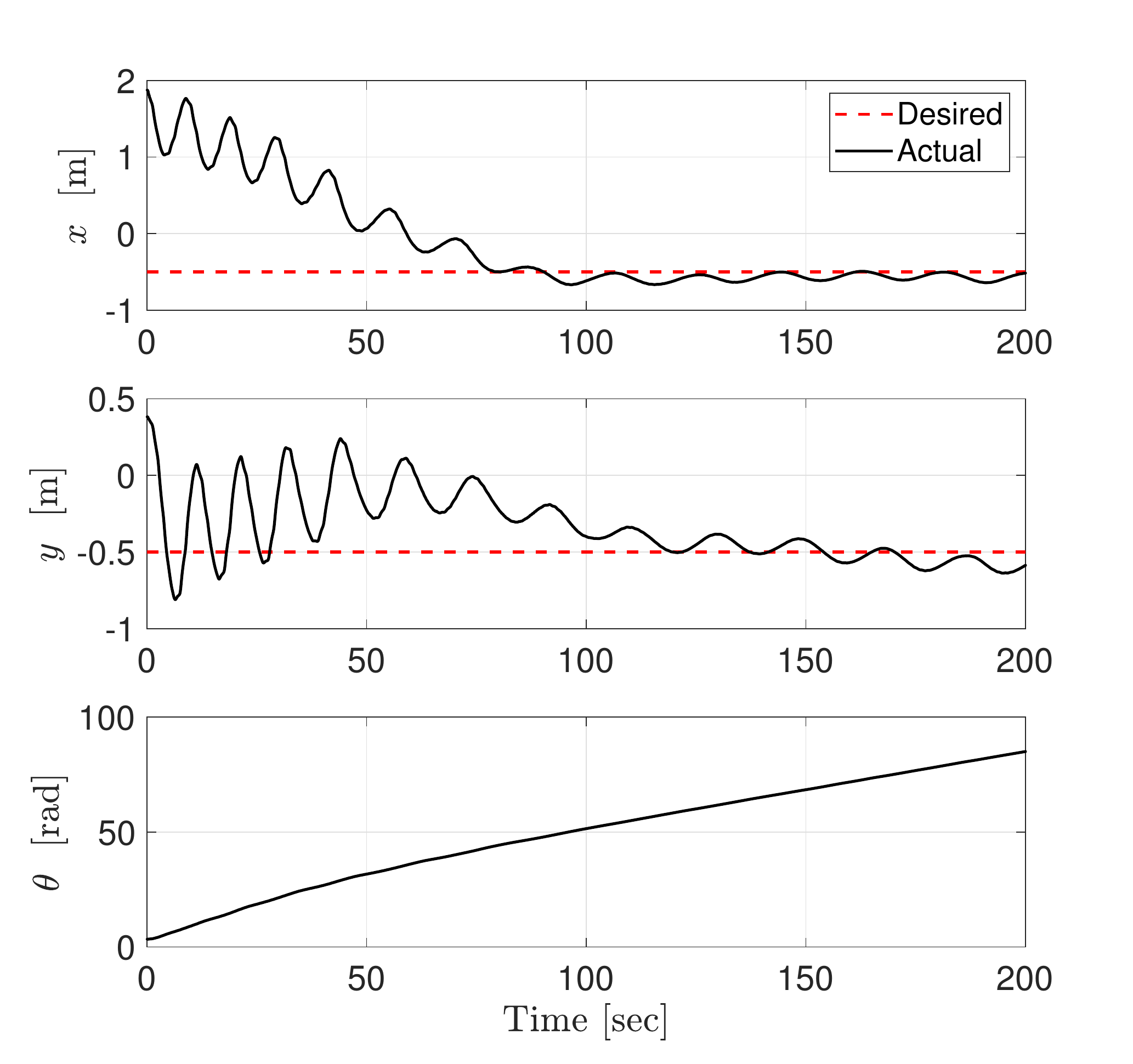}
         \caption{}
         \label{fig:traj-2e}
     \end{subfigure}
        \caption{Experimental path and configuration trajectories of the underactuated boat in source seeking (Case 2: $c=5.5\times 10^{-3},\varepsilon=0.2,k=0.08$).}
        \label{fig:experiments2}
\end{figure}

\begin{figure}[!t]
	\centering
	\includegraphics[scale=0.325]{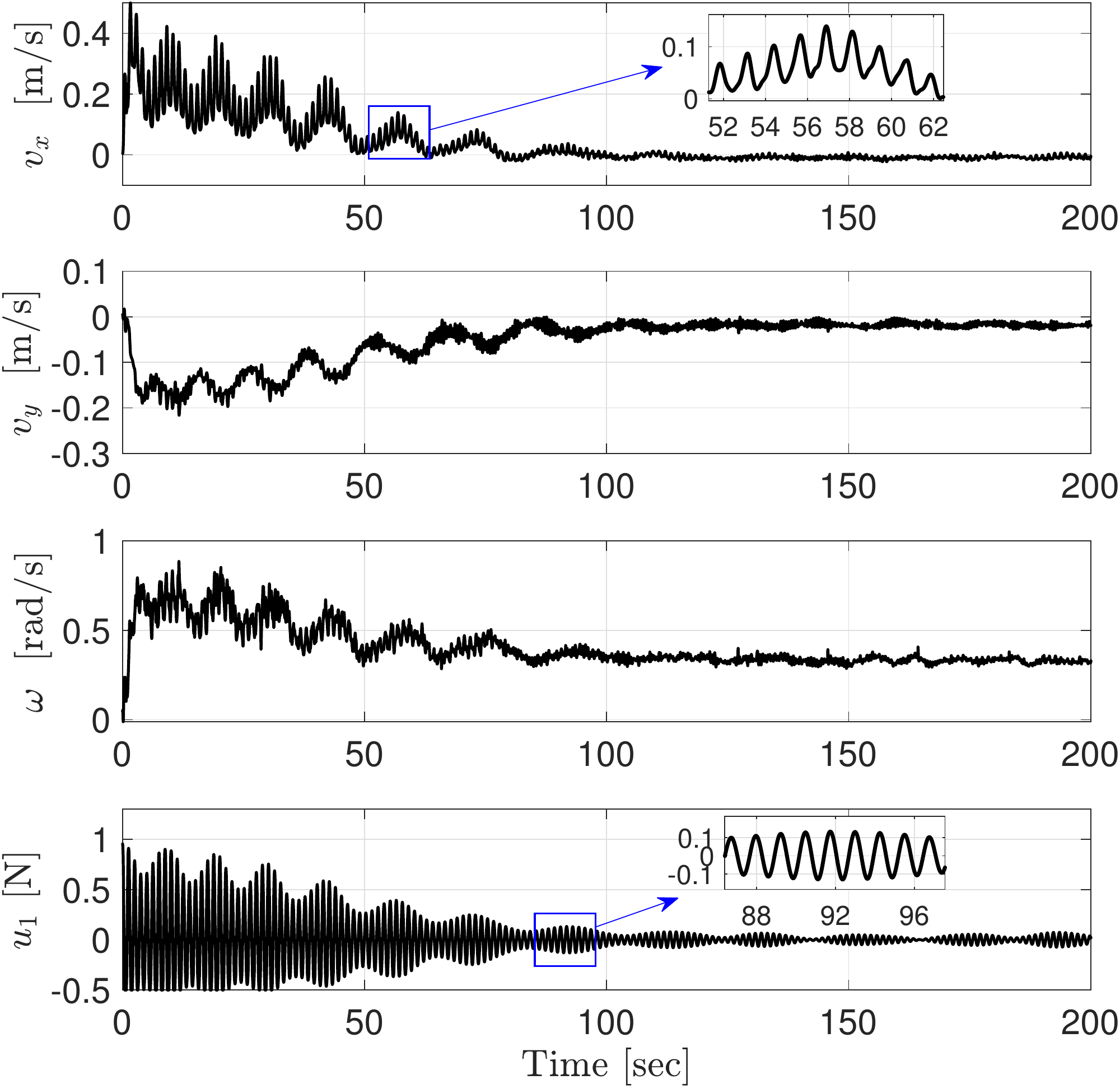}
	\caption{Experimental velocity trajectories and the surge force of the underactuated boat in source seeking (Case 2: $(c=5.5\times 10^{-3},\varepsilon=0.2,k=0.08$).}
	\label{fig:8d}
\end{figure}

\begin{figure}[!t]
     \centering
     \begin{subfigure}[b]{0.48\textwidth}
         \centering
         \includegraphics[scale=0.325]{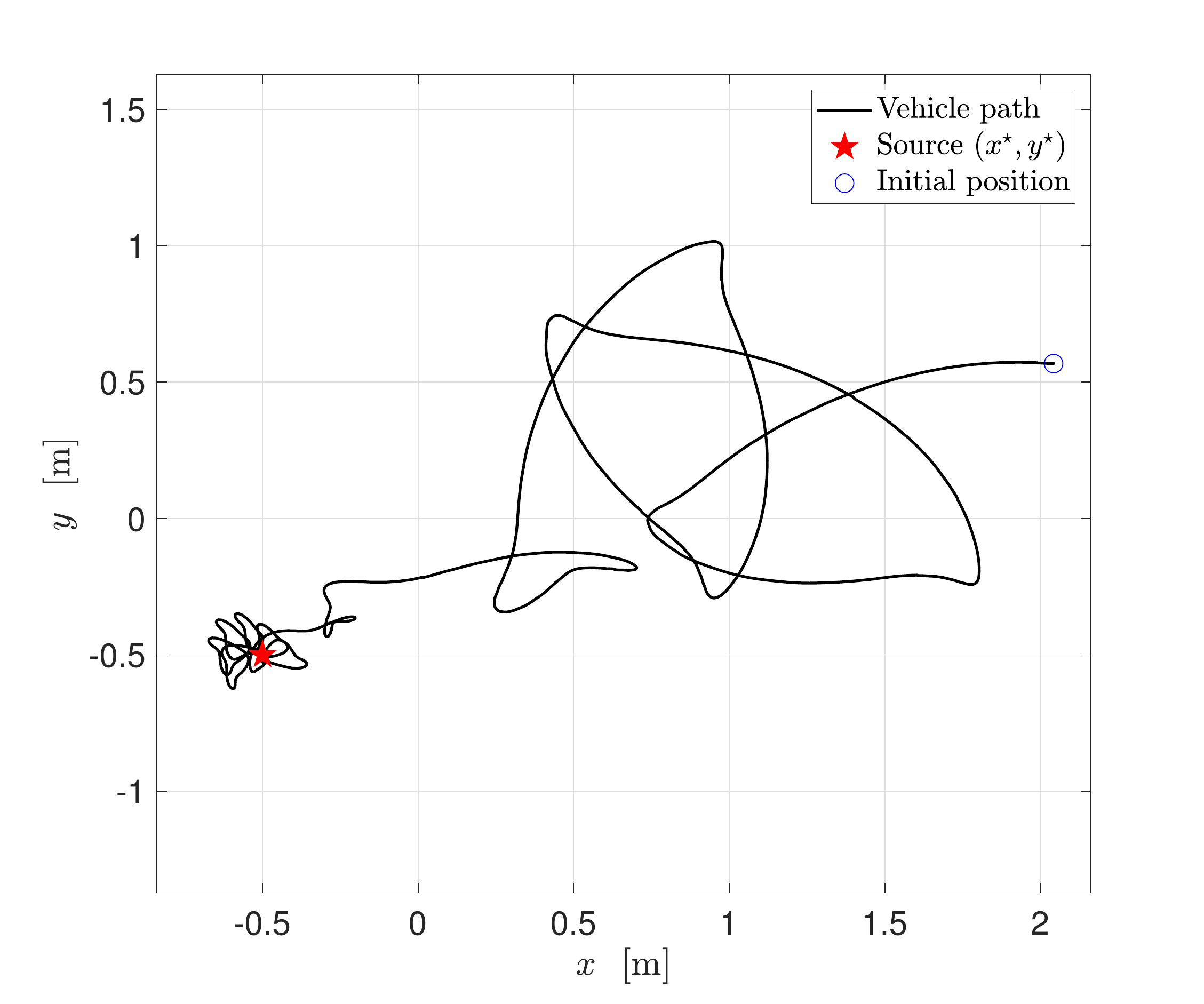}
         \caption{}
         \label{fig:path-3e}
     \end{subfigure}
     \hfill
     \begin{subfigure}[b]{0.48\textwidth}
         \centering
         \includegraphics[scale=0.325]{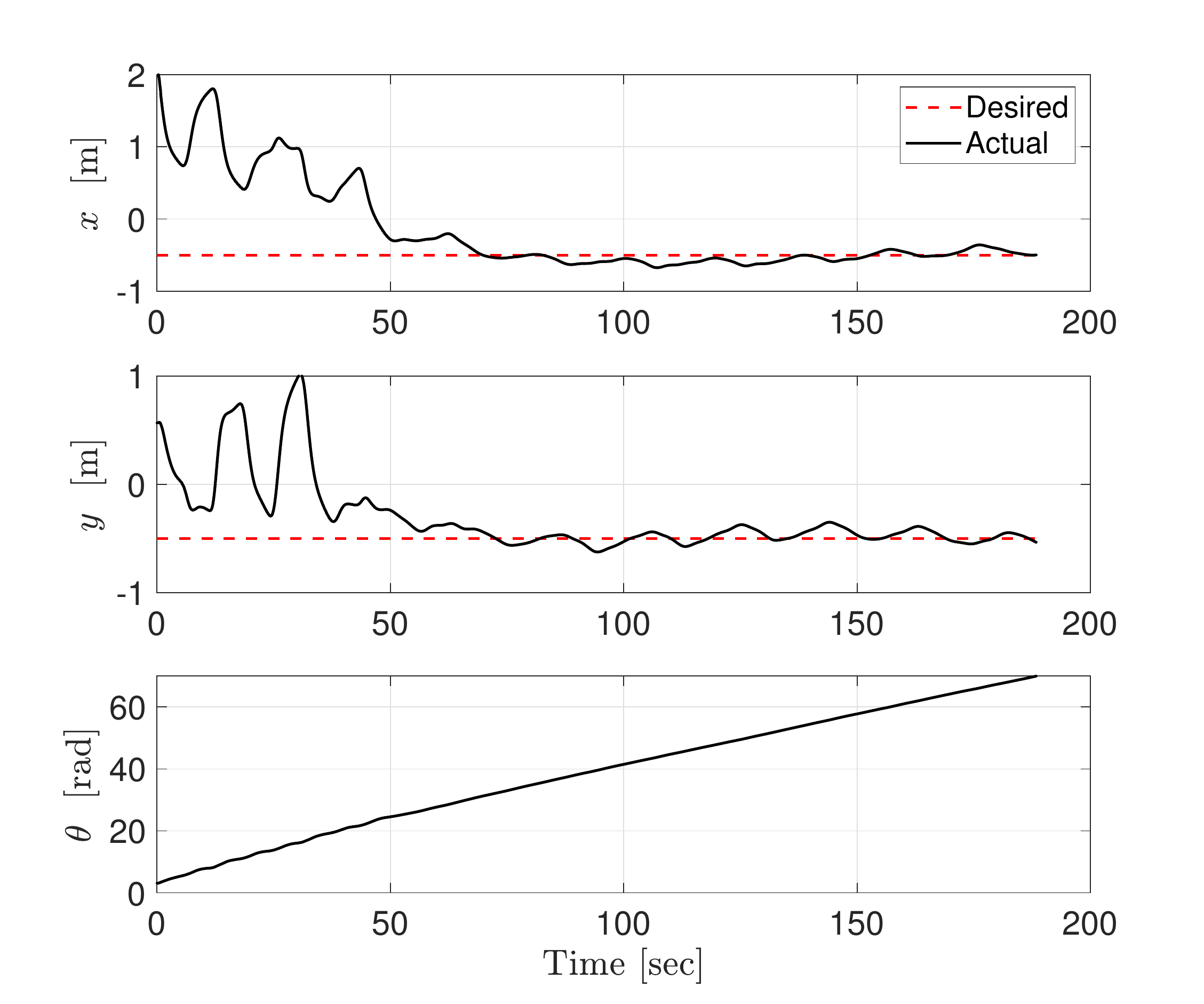}
         \caption{}
         \label{fig:traj-3e}
     \end{subfigure}
        \caption{Experimental path and configuration trajectories of the underactuated boat in source seeking (Case 3: $c=5.5\times 10^{-3},\varepsilon=1,k=0.33$).}
        \label{fig:experiments3}
\end{figure}

\begin{figure}[!t]
	\centering
	\includegraphics[scale=0.325]{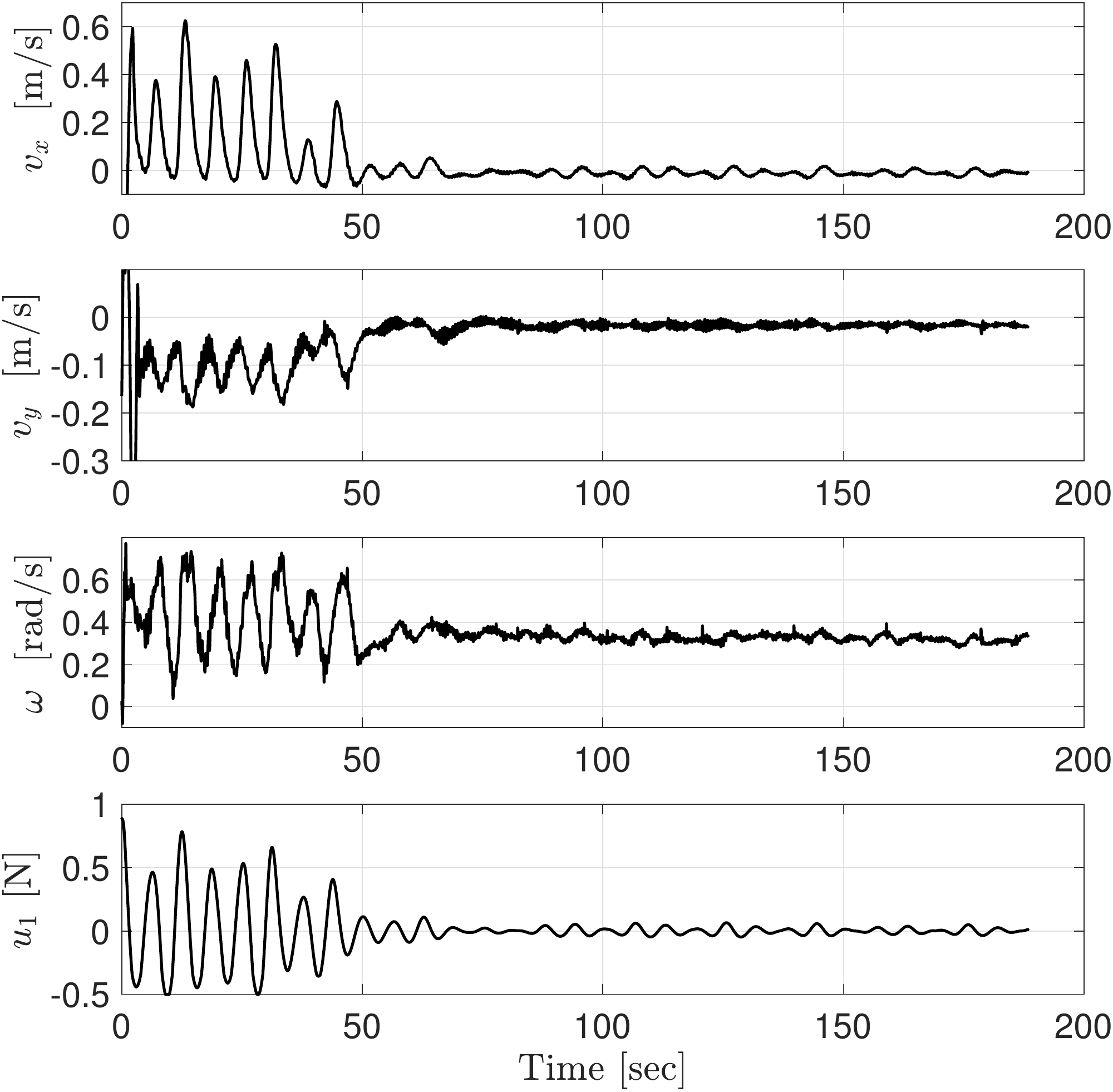}
	\caption{Experimental velocity trajectories and the surge force of the underactuated boat in source seeking (Case 3: $c=5.5\times 10^{-3},\varepsilon=1,k=0.33$).}
	\label{fig:7d}
\end{figure}

In the experiments, we put the feature point at the position $(x^\star,y^\star)=(-0.5,-0.5)$~m and the boat starts from rest. In the first case, we selected the control parameters as $c=2.2\times 10^{-3}$, $\varepsilon=0.5$, and $k=0.08$. Figure \ref{fig:experiments1} shows the boat's path and pose time history and demonstrates that the boat successfully finds the source. Figure \ref{fig:9d} shows the boat's velocity and surge force time history. 
In the second case, we increased the torque and frequency by letting $c=5.5\times 10^{-3}$ and $\varepsilon=0.2$, and kept $k=0.08$. Figure \ref{fig:experiments2} shows that the boat speeds up and goes through a higher frequency motion. As a result, the control effort is much higher to travel a larger distance without any improvement in convergence speed. Figure \ref{fig:8d} shows the boat's velocity and surge force time history in the second case. 
In the third case, we kept the higher torque value $c=5.5\times 10^{-3}$ but increased $\varepsilon=1$ and $k=0.33$. It can be seen from Fig. \ref{fig:experiments3} that increasing the gain $k$ may improve convergence speed, but on the other hand, the path becomes less predictable. Figure \ref{fig:7d} shows the boat's velocity and surge force time history in the third case. It can be observed from Figs. \ref{fig:9d}, \ref{fig:8d}, and \ref{fig:7d} that the boat was driven by surge force that fluctuates around zero. However, the surge velocity does not necessarily fluctuate around zero because the backward motion of the boat has a larger damping force compared with the forward motion.

It can be seen from both the simulations and the experiments that the larger the parameter $c$, the faster the angular motion; the larger the parameter $k$, the faster the convergence speed; and the larger the parameter $\varepsilon$, the higher the frequency of the back-and-forth motion, but the higher the precision of the convergence. Note that the observed winding motion is typical in ES due to the use of only one signal for feedback.


\section{Conclusions}\label{sec:conclusions}
The ES design for force-controlled underactuated mechanical systems without position or velocity measurements was previously an open problem. In this work, we developed a source seeking scheme for generic force-controlled strictly dissipative planar underactuated vehicles by surge force tuning. The control design is based on symmetric product approximations, averaging, passivity, and partial-state stability theory. The controller does not require any position or velocity measurements but only real-time measurements of the source signal at the current position. The P-SPUAS is proven for the closed-loop source seeking system. Both numerical simulations and experimental results of an underactuated boat are presented to illustrate the performance of the proposed source seeker. Our future research will focus on the extension of the presented approach to multi-agent source seeking \cite{sahal2019switching}, and source seeking for underactuated vehicles by torque tuning.

\appendices
\section{The Variation of Constants Formula}\label{appendix:geometric}
Consider the dynamic system 
\begin{equation}\label{eqn:ODE}
    \dot{x}=g(t,x), \quad x(0)=x_0,
\end{equation}
where the vector field $g(t,x)$ is locally Lipschitz in $x$ uniformly in $t$. The \textit{flow map} $\Phi_{0,\,t}^g(\cdot)$ is a diffeomorphism, which describes the solution of (\ref{eqn:ODE}) at time $t$, i.e., $x(t)=\Phi_{0,\,t}^g(x_0)$.

Given a diffeomorphism $\phi$ and a vector field $f$, the \textit{pull back} of $f$ along $\phi$, denoted by $\phi^*f$, is the vector field
\begin{equation}
    (\phi^*f)(x)\coloneqq\left(\frac{\partial \phi^{-1}}{\partial x}\circ f\circ \phi\right)(x),
\end{equation}
where $(f\circ\phi)(x)=f\left(\phi(x)\right)$. The variation of constants formula \cite{bullo2002averaging,bullo2005geometric} characterizes the relationship between the flow of $f+g$ and the flows of $f$ and $g$.

\begin{theorem}[Variation of constants formula]\label{thm:variation-of-constants}
    Consider the dynamic system 
    \begin{equation}\label{eqn:voc}
        \dot{x}=f(t,x)+g(t,x),\quad x(0)=x_0,
    \end{equation}
    where $f,g:\mathbb{R}_{\ge 0}\times\mathbb{R}^n\to\mathbb{R}^n$ are smooth vector fields. If $z(t)$ is the solution of the system
    \begin{equation}\label{eqn:pull-back}
        \dot{z}(t)=\left(\left(\Phi_{0,\,t}^g\right)^*f\right)(t,z), \quad z(0)=x_0,
    \end{equation}
    then the solution $x(t)$ of the initial value problem
    \begin{equation}
        \dot{x}=g(t,x),\quad x(0)=z(t)
    \end{equation}
    is the solution of system (\ref{eqn:voc}). 
\end{theorem}

System (\ref{eqn:pull-back}) is called the \textit{pull back system}. Furthermore, if $f$ is a time-invariant vector field and $g$ is a time-varying vector field, then the pull back of $f$ along $\Phi_{0,\,t}^g$ is given by
\begin{equation}
\begin{split}
    &\left(\left(\Phi_{0,\,t}^g\right)^*f\right)(t,x)=f(x)\\
    & +\sum_{k=1}^\infty \int_0^t\cdots\int_0^{s_{k-1}}\left({\rm ad}_{g(s_k,x)}\cdots{\rm ad}_{g(s_1,x)}f(x)\right) {\rm d}s_k\cdots{\rm d}s_1.
\end{split}
\end{equation}

\section{Proof of Proposition \ref{thm:partial-SPUAS}}\label{appendix:stability}
We successively prove that conditions 1, 2, and 3 of Definition \ref{def:partial-practical-stability} are satisfied.
\begin{enumerate}[1)]
\item Take an arbitrary $c_2>0$, and let $b_2\in(0,c_2)$. By the P-US property, there exists $c_1$ such that
\begin{equation*}
|x_{10}|\le c_1 \implies  |x_1(t)|\le b_2, \quad \forall t\ge t_0,~ \forall x_{20}\in\mathbb{R}^{n_2}.
\end{equation*}
Let $b_1\in(0,c_1)$, and by the P-UGA property, there exists $T$ such that 
\begin{equation*}
|x_{10}|\le c_1 \implies  |x_1(t)|\le b_1, \quad \forall t\ge t_0+T,~ \forall x_{20}\in\mathbb{R}^{n_2}.
\end{equation*}
Let $d=\min\{c_1-b_1,c_2-b_2\}$ and $K=\{(x_1,x_2)\in\mathbb{R}^{n_1}\times\mathbb{R}^{n_2} :|x_1|\le c_1, |x_2|\le r\}$, where $r>0$ is an arbitrary number. By the partial converging trajectory property, there exists $\varepsilon^*$ such that for all $(x_{10},x_{20})\in K$ and for all $\varepsilon\in(0,\varepsilon^*)$,
\begin{equation*}
    |x_1^\varepsilon(t)-x_1(t)|<d,\quad \forall t\in[t_0,t_0+T].
\end{equation*}
Thus, we conclude that for all $t_0\in\mathbb{R}_{\ge 0}$, for all $(x_{10},x_{20})\in K$ and for all $\varepsilon\in(0,\varepsilon^*)$,
\begin{equation}\label{eqn:pratial-US}
\begin{split}
&|x_1^\varepsilon(t)|<c_2,\quad \forall t\in[t_0,t_0+T],\\
&|x_1^\varepsilon(t)|<c_1,\quad \text{for}~ t=t_0+T.
\end{split}
\end{equation}
Since $|x_1^\varepsilon(t_0+T)|<c_1$, a repeated application of (\ref{eqn:pratial-US}) yields that
for all $(x_{10},x_{20})\in K$ and for all $\varepsilon\in(0,\varepsilon^*)$, we have $|x_1^\varepsilon(t)|<c_2$, $\forall t\ge t_0$.

\item Take an arbitrary $c_1>0$, and let $b_1\in(0,c_1)$. By the P-UGB and P-UGA properties, there exist $b_2$ and $T$ such that for all $t_0\in\mathbb{R}_{\ge 0}$ and for all $x_{20}\in\mathbb{R}^{n_2}$,
\begin{equation*}
\begin{split}
&|x_{10}|\le c_1 \implies |x_1(t)|\le b_2,\quad \forall t\ge t_0,\\
&|x_{10}|\le c_1 \implies |x_1(t)|\le b_1,\quad \forall t\ge t_0+T.
\end{split}
\end{equation*}
Let $c_2>b_2$, and by the partial converging trajectory property again, we conclude that there exists $\varepsilon^*$ such that for all $(x_{10},x_{20})\in K$ and for all $\varepsilon\in(0,\varepsilon^*)$, we have $|x_1^\varepsilon(t)|<c_2$, $\forall t\ge t_0$.

\item Take arbitrary $c_1,c_2>0$. By the Item 1 proven above, there exist $c_3$ and $\varepsilon^*$ such that for all $t_0\in \mathbb{R}_{\ge 0}$, for all $\varepsilon\in(0,\varepsilon^*)$,
\begin{equation}\label{eqn:pratial-US2}
    |x_{10}|\le c_3\implies |x_1^\varepsilon(t)|<c_2, \forall t\ge t_0,~\forall x_{20}\in\bar{\mathcal{B}}_r^{n_2}.
\end{equation}
Let $b_3\in(0,c_3)$, and by the P-UGA property, there exists $T$ such that for all $x_{20}\in\mathbb{R}^{n_2}$,
\begin{equation*}
|x_{10}|\le c_1 \implies  |x_1(t)|\le b_3, \quad \forall t\ge t_0+T.
\end{equation*}
Let $d=c_3-b_3$. Then, by the partial converging trajectory property, there exists $\varepsilon^\#$ such that for all $\varepsilon\in(0,\varepsilon^\#)$ and for all $x_{20}\in\bar{\mathcal{B}}_r^{n_2}$,
\begin{equation*}
|x_{10}|\le c_1\implies |x_1^\varepsilon(t)-x_1(t)|<d,\quad \forall t\in[t_0,t_0+T],
\end{equation*}
which implies that for all $\varepsilon\in(0,\varepsilon^\#)$ and for all $x_{20}\in\bar{\mathcal{B}}_r^{n_2}$,
\begin{equation*}
|x_{10}|\le c_1\implies |x_1^\varepsilon(t_0+T)|<c_3.
\end{equation*}
Finally, together with (\ref{eqn:pratial-US2}), we conclude that for all $t_0\in\mathbb{R}_{\ge 0}$, for all $\varepsilon\in (0,\min\{\varepsilon^*,\varepsilon^\#\})$, and for all $x_{20}\in\bar{\mathcal{B}}_r^{n_2}$, 
\begin{equation*}
|x_{10}|\le c_1\implies |x_1^\varepsilon(t)|<c_2, \quad \forall t\ge t_0+T,
\end{equation*}
which completes the proof.\qed
\end{enumerate}

\bibliographystyle{IEEEtranS}
\bibliography{root}

\begin{IEEEbiography}[{\includegraphics[width=1in,height=1.25in,clip,keepaspectratio]{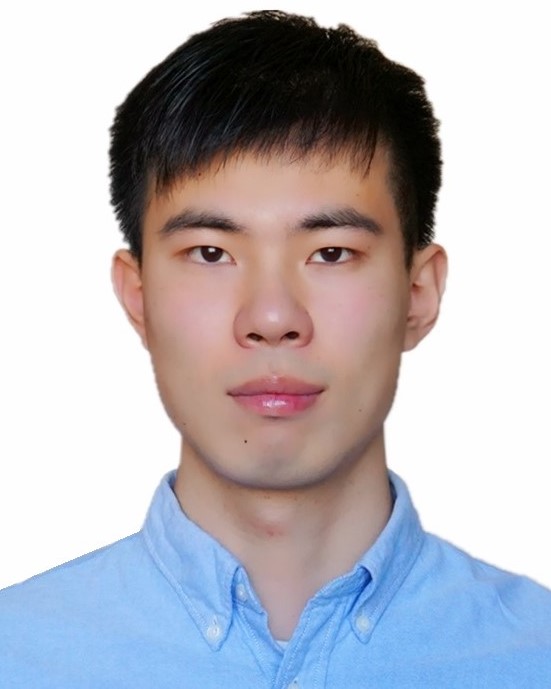}}]{Bo Wang} 
(Graduate Student Member, IEEE) received the M.S. degree in control theory and engineering from University of Chinese Academy of Sciences, Beijing, China, in 2018. He is currently a Ph.D. candidate at the Department of Mechanical Engineering, Villanova University, United States. 

His research interests include nonlinear control theory (robust, adaptive, passive, etc.), underactuated systems, nonholonomic systems, geometric control theory, networked control systems, extremum seeking control, and robotics.
\end{IEEEbiography}

\begin{IEEEbiography}[{\includegraphics[width=1in,height=1.25in,clip,keepaspectratio]{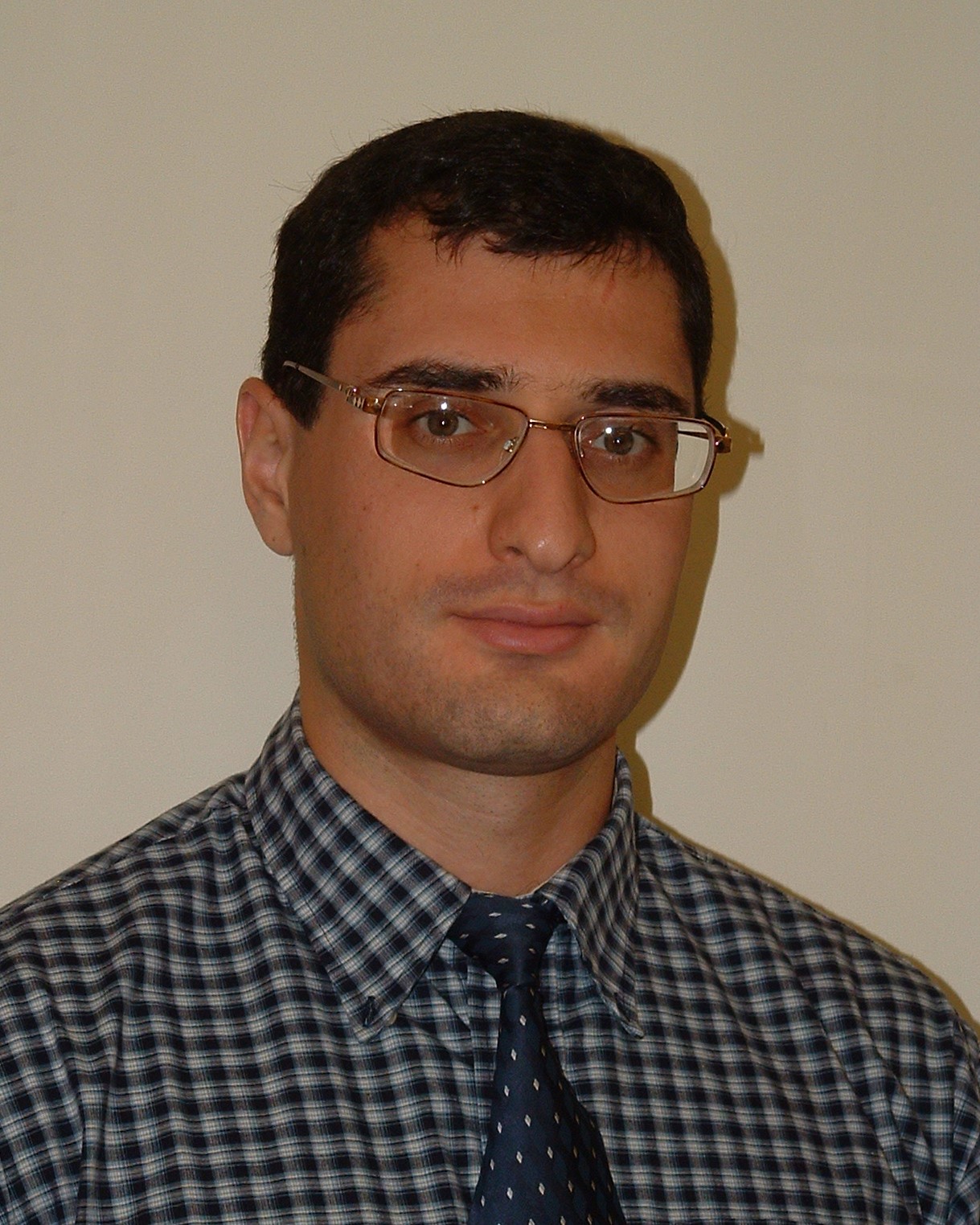}}]{Sergey Nersesov}
(Member, IEEE) received the B.S. and M.S. degrees in aerospace engineering from the Moscow Institute of Physics and Technology, Zhukovsky, Russia, in 1997 and 1999, respectively. In 2003 he received the M.S. degree in applied mathematics and in 2005 he received the Ph.D. degree in aerospace engineering both from the Georgia Institute of Technology, Atlanta, GA. Currently, he is an Associate Professor in the Department of Mechanical Engineering at Villanova University, Villanova, PA. His research interests include nonlinear dynamical system theory, large-scale systems, cooperative control for multi-agent systems, and hybrid and impulsive control for nonlinear systems. He is a coauthor of the books \textit{Thermodynamics: A Dynamical Systems Approach (Princeton University Press, 2005)}, \textit{Impulsive and Hybrid Dynamical Systems: Stability, Dissipativity, and Control (Princeton University Press, 2006)}, and \textit{Large-Scale Dynamical Systems: A Vector Dissipative Systems Approach (Princeton University Press, 2011)}.
\end{IEEEbiography}

\begin{IEEEbiography}[{\includegraphics[width=1in,height=1.25in,clip,keepaspectratio]{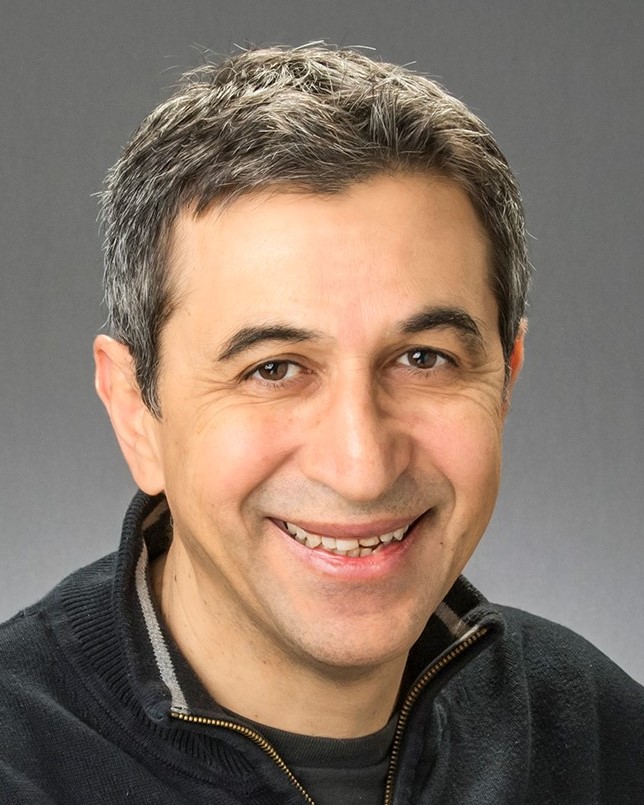}}]{Hashem Ashrafiuon}
(Senior Member, IEEE) received his B.S., M.S. and Ph.D. degrees in Mechanical Engineering from the State University of New York at Buffalo. He joined Villanova University faculty after graduating in 1988. He currently holds the position of Professor in the Department of Mechanical Engineering at Villanova University. He is a fellow of ASME and a senior member of IEEE. He is a senior editor for \textit{Journal of Vibration and Control} and has been on editorial boards of several IEEE and ASME publications. His research interests include nonlinear control of heterogeneous autonomous vehicles and underactuated systems.
\end{IEEEbiography}

\begin{IEEEbiography}[{\includegraphics[width=1in,height=1.25in,clip,keepaspectratio]{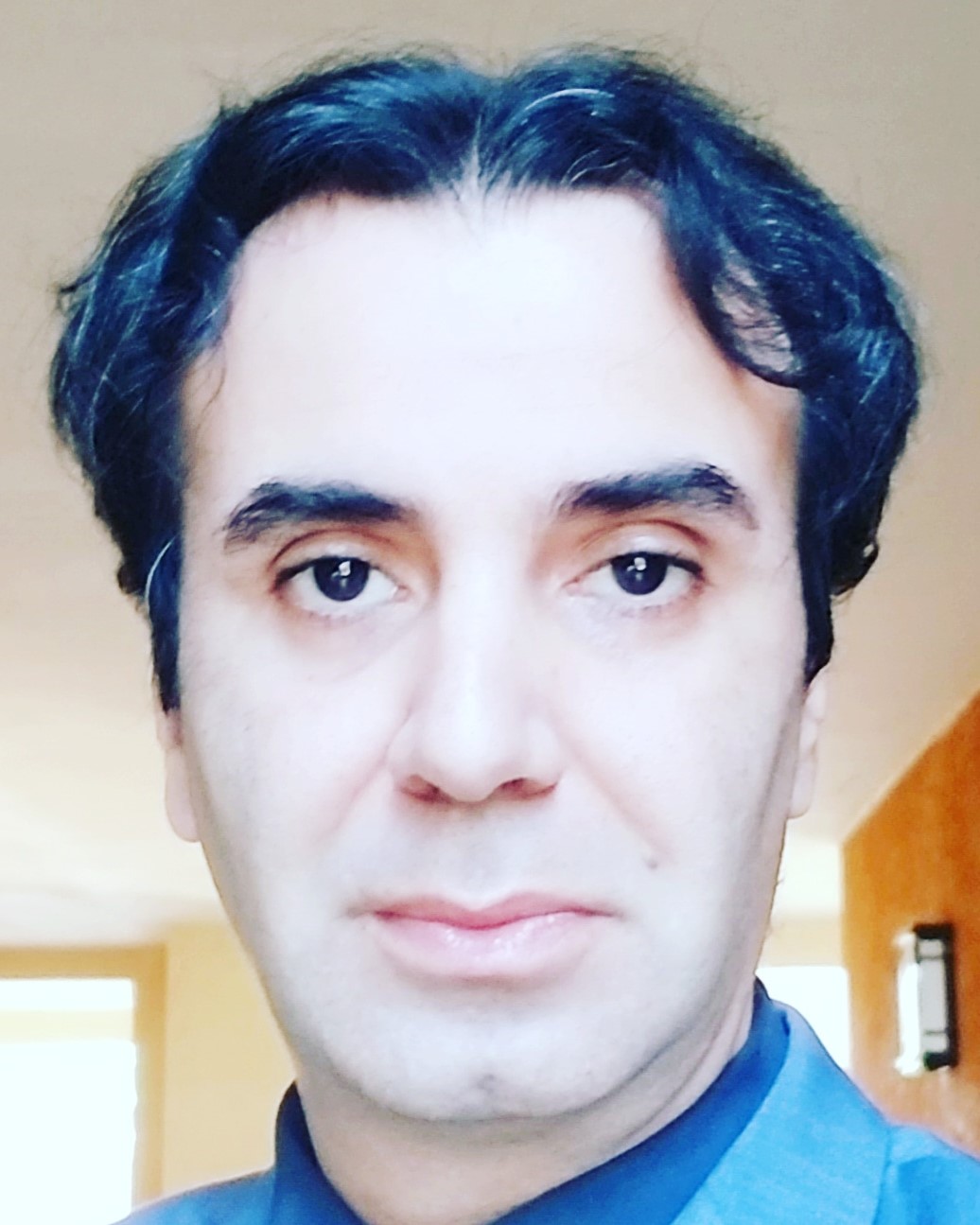}}]{Peiman Naseradinmousavi}
received the B.Sc. degree in mechanical engineering (dynamics and control) from the University of Tabriz, Tabriz, Iran, in 2002, and the Ph.D. degree in mechanical engineering (dynamics and control) from Villanova University, Villanova, PA, USA, in 2012.
He is currently an Associate Professor with the Dynamic Systems and Control Laboratory (DSCL), Department of Mechanical Engineering, San Diego State University, San Diego, CA, USA. His research interests include robotics, smart flow distribution network, nonlinear dynamics, control theory, optimization, magnetic bearings, and mathematical modeling.

Dr. Naseradinmousavi was a recipient of the John J. Gallen Memorial Alumni Award, 2021. He serves as an Associate Editor of \textit{ASME Letters in Dynamic Systems and Control} and the \textit{Journal of Vibration and Control (JVC)}.
\end{IEEEbiography}

\begin{IEEEbiography}[{\includegraphics[width=1in,height=1.25in,clip,keepaspectratio]{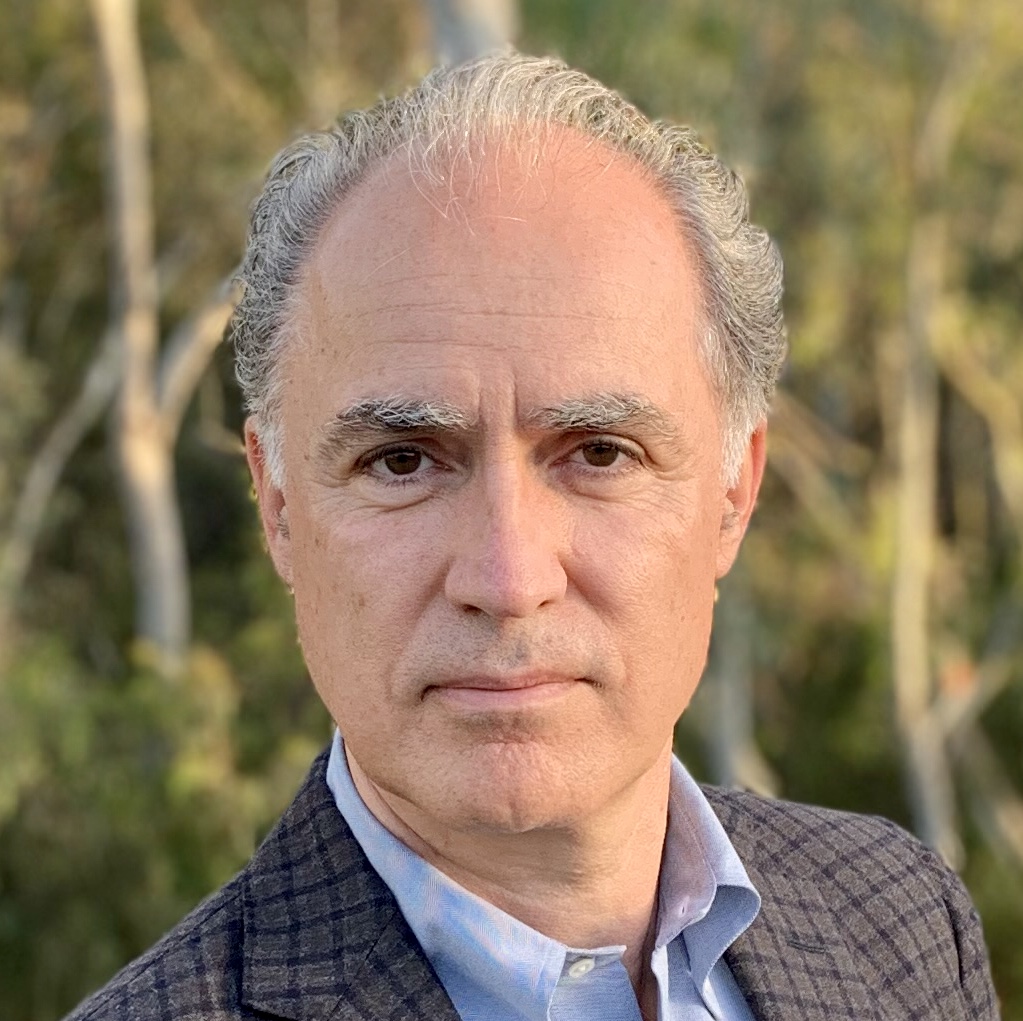}}]{Miroslav Krsti{\'c}}
(Fellow, IEEE) is a Distinguished Professor of mechanical and aerospace engineering, holds the Alspach Endowed Chair, and is the Founding Director of the Cymer Center for Control Systems and Dynamics, University California San Diego (UCSD), La Jolla, CA, USA. He also serves as a Senior Associate Vice Chancellor for Research with UCSD. He has coauthored 16 books on adaptive, nonlinear, and stochastic control, extremum seeking, control of PDE systems including turbulent flows, and control of delay systems.

Krsti{\'c} has been an elected fellow of seven scientific societies-IFAC, ASME, SIAM, AAAS, IET (U.K.), and AIAA (Associate Fellow)-and as a foreign member of the Serbian Academy of Sciences and Arts and of the Academy of Engineering of Serbia. He won the UC Santa Barbara Best Dissertation Award and the Student Best Paper Awards at CDC and ACC,
as a Graduate Student. He has received the Richard E. Bellman Control Heritage Award, SIAM Reid Prize, the ASME Oldenburger Medal, the A.V. Balakrishnan Award for Mathematics of Systems, the Nyquist Lecture Prize, the Paynter Outstanding Investigator Award, the Ragazzini Education Award, the Chestnut textbook prize, Control Systems Society Distinguished Member Award, the PECASE, the NSF Career, the ONR Young Investigator Awards, the Axelby and Schuckpaper prizes, and the first UCSD Research Award given to an Engineer. He has been awarded the Springer Visiting Professorship at UC Berkeley, the Distinguished Visiting Fellowship of the Royal Academy of Engineering, and the Invitation Fellowship of the Japan Society for the Promotion of Science. He serves as an Editor-in-Chief of \textit{Systems and Control Letters} and has been serving as a Senior Editor in \textit{Automatica} and \textit{
IEEE Transactions on Automatic Control}, as an Editor of two Springer book series, and has served as the Vice President for Technical Activities of the IEEE Control Systems Society and as the Chair of the IEEE CSS Fellow Committee.
\end{IEEEbiography}

\enlargethispage{-3in}

\end{document}